\documentclass[sn-mathphys,Numbered, pdflatex,iicol]{sn-jnl}

\usepackage{amsmath,amssymb,amsfonts}%
\usepackage{cleveref}%
\usepackage[title]{appendix}%
\usepackage{manyfoot}%
\usepackage{subcaption}

\usepackage{nameref}

\usepackage{physics}

\DeclareMathOperator{\DCT}{DCT}
\renewcommand{\vec}[1]{\mathbf{#1}}

\renewcommand\appendixname{Supplementary}
\Crefname{appsec}{Supplementary material}{Supplementary materials}

\usepackage{todonotes}

\begin{document}

\title[Article Title]{Spectral-based detection of chromatin loops in multiplexed super-resolution FISH data }

\author*[1,2]{\fnm{Michaël Liefsoens}}\email{michael.liefsoens@kuleuven.be}

\author*[2]{\fnm{Timothy Földes}}
\email{timothy.foldes@sorbonne-universite.fr}

\author[2]{\fnm{Maria Barbi}}
\email{maria.barbi@sorbonne-universite.fr}

\affil*[1]{\orgdiv{Physics department}, \orgname{KU Leuven},  \city{3001 Leuven}, \country{Belgium}}

\affil[2]{\orgdiv{LPTMC}, \orgname{Sorbonne Université, CNRS}, \city{F-75005 Paris}, \country{France}}

\abstract{
Involved in mitotic condensation, interaction of transcriptional regulatory elements or isolation of structural domains, understanding loop formation is becoming a paradigm in the deciphering of chromatin architecture and its functional role.
Despite the emergence of increasingly powerful genome visualization techniques, the high variability in cell populations and the randomness of conformations still make loop detection a challenge.
We introduce a new approach for determining the presence and frequency of loops in a collection of experimental conformations obtained by multiplexed super-resolution imaging. 
Based on a spectral approach, in conjunction with neural networks, this method offers a powerful tool to detect loops in large experimental data sets, both at the population and single cell level.
The method's performance is confirmed by applying it to recently published experimental data, where it provides a detailed and statistically quantified description of the global architecture of the chromosomal region under study. 
}

\keywords{Chromatin architecture, Loops, Multiplexed super-resolution imaging, Spectral density, Neural networks}

\maketitle

\section*{ }\label{sec:introduction}
Loop formation is central to understanding chromatin architecture and its functional role. During mitosis, chromatin adopts a compact structure composed of loops, forming a rod-like configuration \cite{Paulson2021}. SMC (structural maintenance of chromosome) proteins like condensins and cohesins play a pivotal role in organizing these loops \cite{Swedlow2003}.
Recent research reveals that loop formation, mediated by proteins such as CCCTC-Binding factor (CTCF) and cohesin, is also critical in interphase, for gene regulation by facilitating interactions between distant enhancers and promoters in mammals \cite{Rao2014,Karpinska2023}, \textit{Drosophila} \cite{Espinola2021}, and yeast \cite{Costantino2020}. The identification of chromatin loops have become central to unraveling gene regulation complexities and spatial genome organization.
Forthermore, cohesin-dependent loops are involved in the segmentation of interphase chromosomes into topologically associating domains (TADs), defined as sub-Mb self-interacting regions, often delimited by CTCF binding. Depletion of CTCF disrupts both TAD loops and insulation of neighboring TADs \cite{Nora2017}.
Key questions arise regarding loop formation mechanisms, their prevalence, determinants of their position and sizes, and biological functions. 

The loop extrusion mechanism \cite{Alipour2012}, primarily involving SMC family proteins like cohesin (in interphase) and condensin (in metaphase), can explain loop formation. Cohesin and CTCF enable loop extrusion by binding to DNA as dimers, after which they act as motors, sliding in opposite directions and enlarging the loops by pulling along the chromatin fibers \cite{Fudenberg2016}. Looping by SMC complexes is observed in various cell types, including mammalian and bacterial cells \cite{Banigan2020}.
As insulator proteins, CTCF and cohesin regulate chromatin loop stability, probably as a 'dynamic complex' that frequently breaks and reforms throughout the cell cycle \cite{Hansen2017}. 

Visualizing the dynamics of loop extrusion in single living cells remains challenging. Fluorescence microscopy tracking two loop anchors has been explored \cite{Gabriele2022, Mach2022}, but requires prior anchor position knowledge and thus a strategy to identify the loops. In high-throughput genomic techniques like Hi-C \cite{Lieberman-Aiden2009}, stable loops manifest as discrete points in contact maps. Data analysis tools detecting DNA loops in contact maps, based on contact count enrichment or specific patterns, are available \cite{Roayaei2020,Salameh2020,Matthey-Doret2020}.

However, Hi-C methods lack the ability to reconstruct the polymer's spatial trajectory, only quantifying contact frequencies between monomers. These limitations might, however, be overcome by combining fluorescence in situ hybridization (FISH) and super-resolution microscopy to achieve high-resolution imaging of individual genomic regions (Hi-M \cite{CardozoGizzi2019}, ORCA \cite{Mateo2019}, OligoFISSEQ \cite{Nguyen2020}, MERFISH \cite{Bintu2018,Su2020}).
These are high-throughput, high-resolution, microscopy-based technologies that, for the first time, allow the visualization of the spatial trajectory of the polymer by sequential labeling and imaging multiple loci along a single chromosome region, in fixed cells.
This results in collections of configurations, sampled with a resolution up to 30 kb, which are for the moment difficult to fully exploit, especially to the aim of loop determination. The most frequent approaches are based on the reconstruction of \textit{distance} maps, then interpreted as contact maps \cite{Lee2023}. 
However, this approach restricts the information to a level already obtainable with previous techniques.

Innovative methods are clearly needed to fully exploit this new data. In this study, we address the possibility of characterizing chromatin loops through a spectral representation of chain configurations, thereby leveraging the whole information of chain 3D spatial arrangement offered by sequential FISH methods.

\section*{Results}\label{sec:results}

\subsection*{Power spectral density reveals large-scale polymer features}

Loops represent a distinctive aspect of chromosome folding, which must be considered within the broader context of the stochastic chromatin architecture. At a macroscopic level, heterochromatin is denser and transcriptionally repressed, while euchromatin is lighter and active, akin to polymers adopting globule and coil conformations, respectively \cite{Mirny2011, Grosberg2012, Barbieri2012, Boettiger2016, Szabo2018, Lesage2019, Foldes2021}. More specifically, super-resolution imaging of epigenetic domains in \textit{Drosophila} \cite{Boettiger2016} seems to indicate that their structure is compatible with the behavior of a {self-attracting polymer} close to the coil-globule transition \cite{Lesage2019}.
This transition, governed by the monomer-monomer interaction parameter\footnote{The effective monomer-monomer interaction parameter $\varepsilon$ depends on factors such as temperature, solvent properties, polymer composition, and external forces.} $\varepsilon$, manifests through state dependent scaling properties of the mean radius of gyration $\langle R_g\rangle$  (or equivalently the end-to-end distance $\langle R\rangle $) as a function of monomer number $N$. Scaling laws, thus, enable the identification of the folding state. Nevertheless, this method necessitates the comparison of polymers with varying lengths, which may not always be feasible.

In prior work \cite{Foldes2021}, we developed an innovative approach for analyzing fluorescent imaging data that overcomes this obstacle. Our method employs spectral analysis of configurations, focusing on long-distance features. Specifically, we apply a discrete cosine transform (DCT) to spatial coordinates and, by taking the mean squared amplitudes of the DCT coefficients $\vec{x}_p$,  we construct a power spectral density (PSD), $\langle \vec{x}_p^2 \rangle$. 
For low $\varepsilon$, in the coil state, PSDs follow the expected scaling $\langle \vec{x}_p^2 \rangle \propto p^{-(1+2\nu)}$, where the exponent $\nu \approx 0. 588$ is the Flory~\cite{Grosberg1994} exponent: this scaling law is indeed the spectral counterpart of Flory's scaling, $\langle R^2\rangle \sim N^{2\nu}$.
However, as $\varepsilon$ increases above a critical value $\varepsilon_\theta(N)$,
the strong attraction induces a second-order phase transition to curled up conformations, called globules \cite{Grassberger1995,Vogel2007, Foldes2021}. Globules have a roughly spherical volume and uniform density, yielding the typical scaling $\langle R^2\rangle \propto N^{2/3}$. Now, this state has a characteristic spectrum that becomes constant for the smallest $p$ modes, making it possible to use the PSD to characterize the coil-globule state of a polymer by identifying its low $p$ spectral scaling \cite{Foldes2021}.

These findings emphasize the significance of examining large scales features, namely the first spectral modes, when probing overall polymer organization. They motivate further exploration to determine if this spectral approach can detect loops in chromosomal regions.

\subsection*{Power spectral density differentiates between looped and non-looped fBm-based polymer models} 

As a first step, we extend the PSD analysis to circular polymers, to examine the impact of looping on the spectrum.
We employ a minimal, yet instructive, model of polymer configurations represented as 3D correlated random walks $\boldsymbol{\gamma}_n$, using fractional Brownian motion (fBm).
The degree of correlation of the fBm is determined by the Hurst exponent $H$:  
\begin{equation}
 C_{\gamma\gamma}(i,j)=\tfrac{1}{2} \sigma_\gamma^2 (i^{2H}+j^{2H}-|i-j|^{2H})
\end{equation}
where $\sigma_\gamma^2 = \langle \gamma_1^2 \rangle$ is the variance of the first step. 
For our theoretical description, we consider polymer conformations with a uniform Hurst exponent $H$. Following \cite{Gasbarra2007} and as detailed in \Cref{Appendix:loopedPSD}, we define a looped fBm as 
\begin{equation}
\label{eq:loopedfBm}
    \boldsymbol{\lambda}_n =\boldsymbol{\gamma}_n - {\cal B}^{(H)}_{n}\, \vec{R}:
\end{equation}
here, $\vec{R}=\boldsymbol{\gamma}_N-\boldsymbol{\gamma}_1$ represents the fBm end-to-end vector and ${\cal B}^{(H)}_{n} 
={N^{-2H}}\, C_{\gamma\gamma}(n,N)$ 
is the appropriate bridge function needed to connect the two ends of the fBm to construct an fBm loop.

For our simplified fBm model, the PSD of the looped chain can be obtained analytically.
Thanks to the linearity of the DCT, the difference between looped and linear fBm is indeed simply the DCT of the bridge function ${\cal B}_H \, \vec{R}$. The symmetry properties of this function then ensure that \textit{(i)} the even modes for looped fBm remain asymptotically unchanged compared to those of the corresponding non-looped; and \textit{(ii)} the odd modes systematically decrease, with the extent of reduction diminishing as the mode number $p$ increases.
These results are proven in \Cref{Appendix:loopedPSD}, 
Additionally, we demonstrate that the latter property is a general consequence of the condition that the first and last monomer coincide, and thus applies to any looped conformation.

\begin{figure*}[pht]
     \centering
     \hfill
     \begin{subfigure}{.3\linewidth}
       \centering
      \includegraphics[width=\linewidth]{{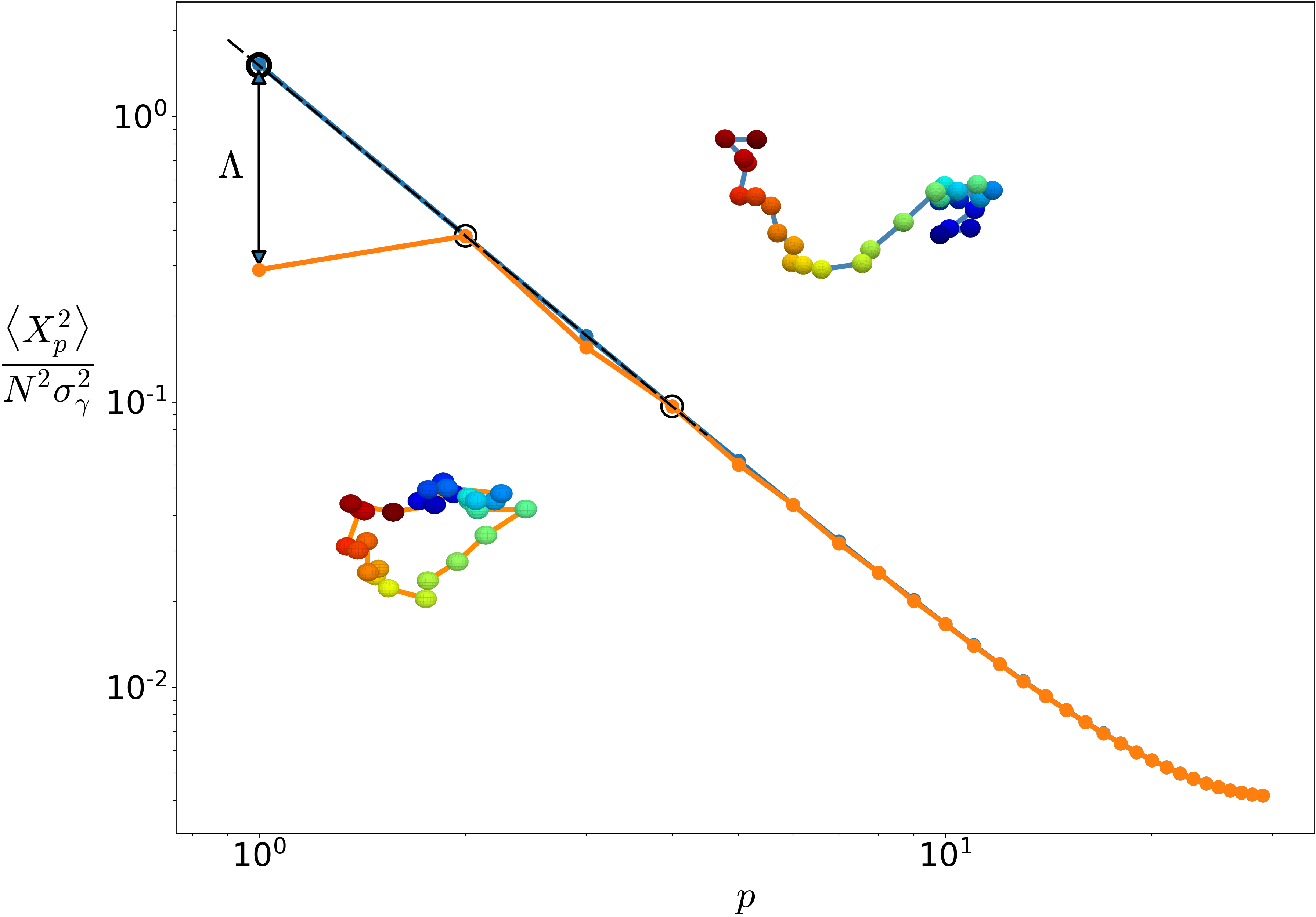}}
      \caption{}
      \label{fig:lambda_def}
    \end{subfigure}\hfill%
    \begin{subfigure}{.3\linewidth}
      \centering
      \includegraphics[width=\linewidth]{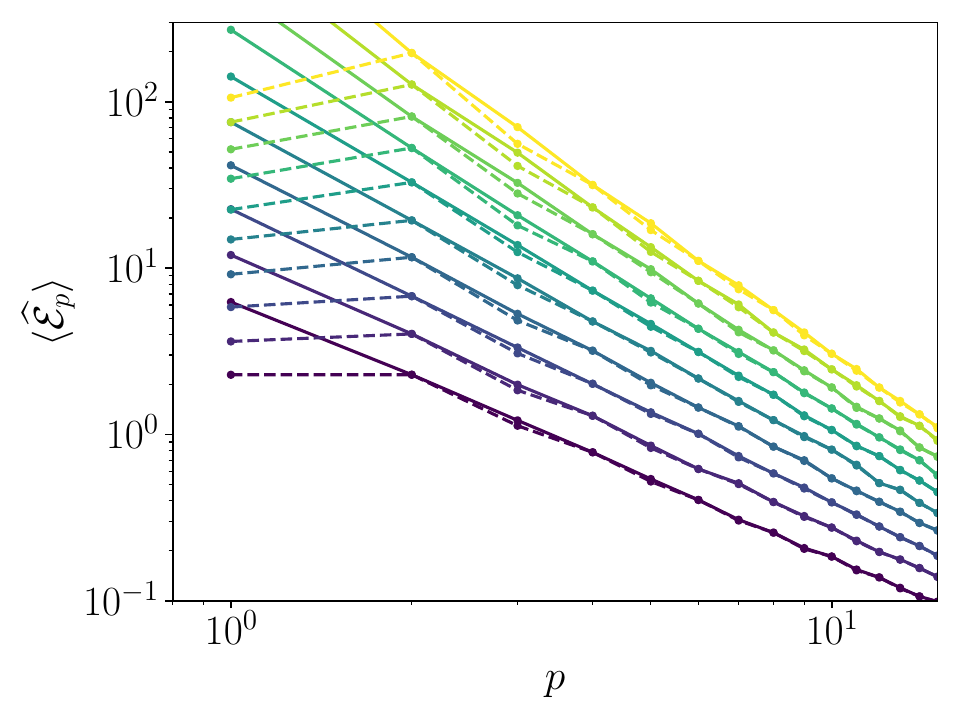}
        \caption{}
        \label{fig:lambda_spectra_varying_H}
    \end{subfigure}\hfill%
    \begin{subfigure}{.3 \linewidth}
      \centering
      \includegraphics[width=\linewidth]{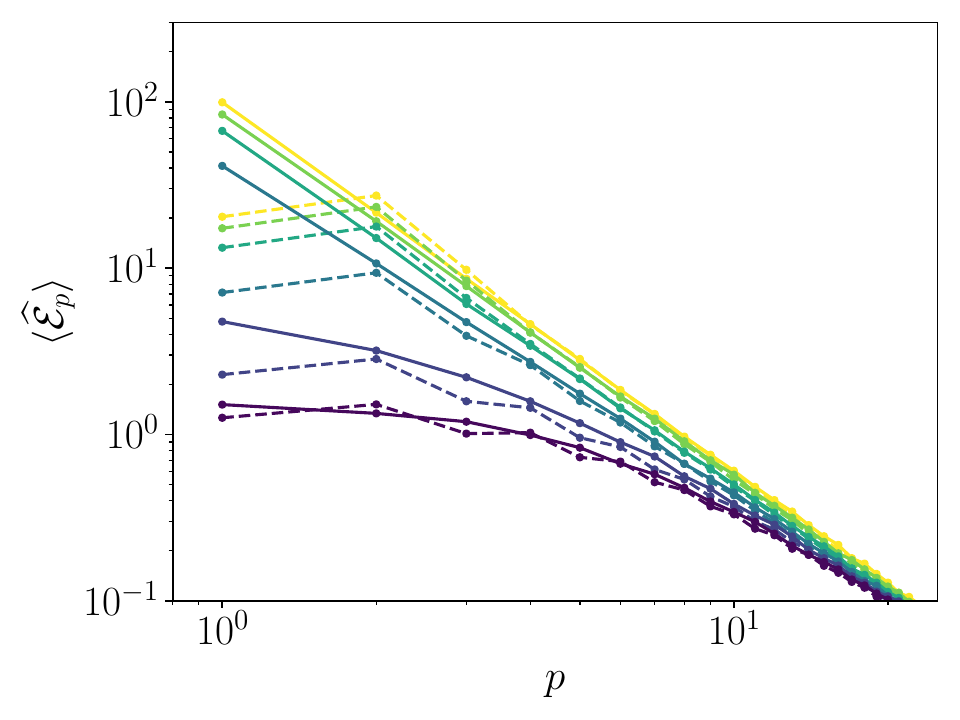}
        \caption{}
        \label{fig:lambda_spectra_varying_eps}
    \end{subfigure}\hfill
    \begin{subfigure}{.95\linewidth}
        \centering
        \includegraphics[width=.8\linewidth]{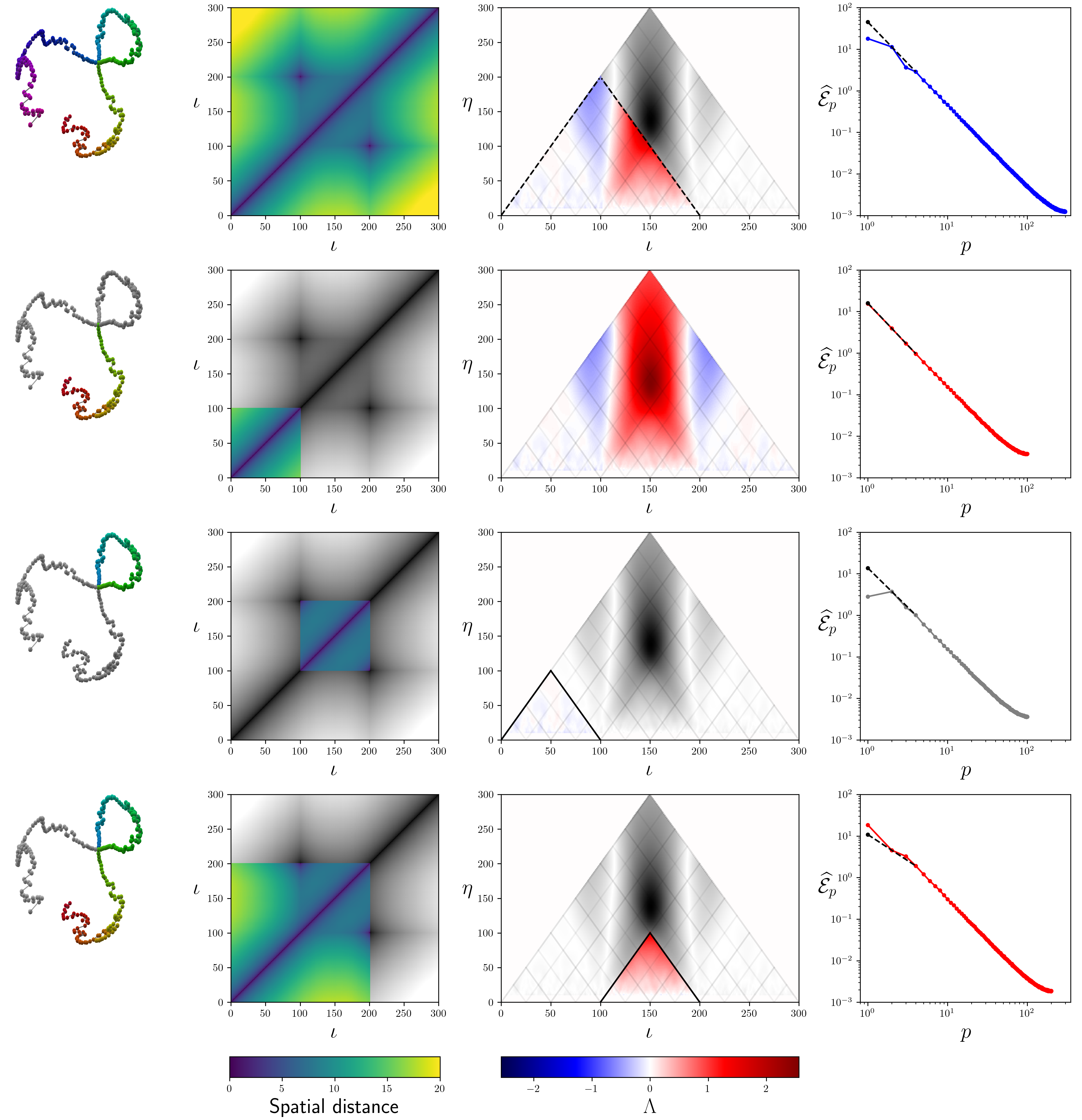}
        \caption{}
        \label{fig:Lambdaplot}
    \end{subfigure}
    \captionof{figure}{\textbf{(a)}
     Theoretical PSD for an $H=0.5$ fBm $\vec{\gamma}_n$ (\Cref{eq:RW-PSD}, blue curve) and the corresponding looped $\vec{\lambda}_n$ (\Cref{eq:LoopRW-PSD}, orange curve). Snapshots show one specific conformation before (upper) and after (lower) looping by means of \Cref{eq:loopedfBm}. $\Lambda(\vec{x})$ is the difference between the observed first mode (here for the looped conformations) and the expected first mode extrapolated from the second and fourth modes (see black dotted line and circles). 
     \textbf{(b)} Estimated PSD for looped and non-looped fBm signals with varying Hurst exponents ($H = 0.3, 0.35, \dots 0.75$), each from samples of $2000$ signals of length $N = 512$.
     \textbf{(c)} Estimated PSD of self-interacting looped and non-looped polymers for $\varepsilon = 0, 0.1, 0.2, 0.3, 0.4, 0.49$. Each spectrum is obtained from samples of $20000$ equilibrium conformations for a polymer size $N = 512$, simulated by the on-lattice Monte Carlo approach.
     \textbf{(d)} $\Lambda$-plot for an ensemble of $2000$ samples of a (random walk) polymer with $N=300$ monomers, all containing an internal loop of size $100$ in the middle (from index 100 to 200). Different rows focus on distinct sub-regions of the same polymer: whole polymer; first third; inner loop; first two thirds (including the loop). The first column displays a \textit{mean} polymer configuration, obtained as described in \nameref{sec:methods}. Sub-regions are colored accordingly. The second column shows the distance map of the polymer, where coloring focus on the selected region.
     The third column shows the $\Lambda$-plot for this polymer ensemble, with colored triangles highlighting regions corresponding to the selected sub-chain. 
     The spectrum for the selected sub-region is shown in column 4.}
\end{figure*}

It's interesting to observe that the difference between looped and non-looped configurations primarily impacts the first modes, emphasizing the pivotal role of large-scale features in defining polymer structure.
The behavior of the PSD for non-looped and looped fBm polymer configurations, is visually depicted in \Cref{fig:lambda_def}.

\subsection*{Log-spectral ratio $\Lambda(\vec{x})$ as an effective observable for loops in fBm signals}

These spectral features offer a method to distinguish between looped and non-looped configurations. Consider indeed a statistical ensemble of 3D signal realizations ${\vec{x}_n}$. We introduce the log-spectral ratio $\Lambda(\vec{x}_n)$ for $\vec{x}_n$, defined as the (logarithmic) difference between the observed amplitude of the first mode and the amplitude  and the amplitude predicted on the basis of a power-law extrapolation from the second and fourth modes. Some manipulation (detailed in \Cref{Appendix:Lambda}) yields the following expression for the log-spectral ratio:
\begin{equation}
    \Lambda(\vec{x}_n) = \log \left(  \frac{\langle \vec{x}_2^2\rangle^2}{\langle \vec{x}_1^2\rangle \langle \vec{x}_4^2\rangle} \right).
    \label{eq:Lambda_def}
\end{equation}
\Cref{fig:lambda_def} (a) provides an illustration of this definition.
Based on our fBm model, we can demonstrate that the log-spectral ratio for a non-looped random walk scales as $N^{-2}$ for $N\to \infty$. In contrast, for a looped fBm, it converges to a finite limit of approximately $1.66$, which clearly distinguishes the two configurations (see \Cref{Appendix:Lambda}). 

For finite chains, we determine a discrimination metric by computing the absolute difference between the spectra of looped and non-looped random walks, and then normalizing it by the same difference at infinity ($= 1.66$). This results in a discriminability level ranging between 0 and 1. This quantity can be calculated analytically and converges extremely fast: having $N>6$ is sufficient to achieve a 90 percent discriminability level; $N>20$ guarantees a 99 percent discriminability level.

To ensure the robustness and applicability of the $\Lambda(\vec{x}_n)$ definition for signals with varying degrees of correlation, we calculate and display in \Cref{fig:lambda_spectra_varying_H} the PSD of fBm signals with different Hurst exponents $H$. Clearly, the behavior theoretically described above and shown in \Cref{fig:lambda_def} is always observed, regardless of the value of $H$.

The log-spectral ratio $\Lambda(\vec{x}_n)$ proves therefore to be a robust observable that allows us to determine whether a polymer is in a linear or looped configuration, independently of its degree of correlation. However, our aim is to investigate the presence of loops in chromosomes.  This 
implies two additional issues, which will be addressed in the following sections.
First, as discussed in the introduction, chromatin domains are expected to be near the coil-globule transition \cite{Lesage2019} and exhibit more or less collapsed, globule-like conformations, depending on epigenetics and transcription activity \cite{Boettiger2016, Szabo2018}. Therefore, it is crucial to verify whether the log-spectral ratio remains reliable across the coil-globule transition.
Second, chromatin loops can vary in size and position along the chromosome. Consequently, we need to adapt our approach to this more general case.

\subsection*{$\Lambda$ detects loops across the simulated coil-globule transition}

To validate the log-spectral ratio approach for identifying loop structures in polymeric molecules, regardless of their state along the coil-globule transition, we performed Monte Carlo simulations of a cubic lattice self-avoiding walk with an energy gain of $-\varepsilon$ (in units of $k_BT$) associated with nearest-neighbor "contact", simulating monomer-monomer effective attraction.
Linear and circular polymers were simulated separately, with reptation moves in the former case \cite{Lesage2019} and Crankshaft rotation, wedge flip, and kink-translocation techniques \cite{Vettorel2009} in the latter, which enhanced simulation efficiency. For the circular polymer, the initial configuration was obtained by the \textit{growing SAW's} algorithm outlined in Ref. \cite{Vettorel2009}.

Spectra were then estimated and compared for linear and looped polymers across a range of $\varepsilon$ values from $0$ to $0.5$. As shown in \Cref{fig:lambda_spectra_varying_eps}, the difference between the looped and non-looped configurations of the simulated polymers reproduces the expected behavior. Consequently, $\Lambda(\vec{x})$ can be defined and used in the same way as theoretically predicted.

\subsection*{Efficient loop identification with the $\Lambda$-plot}

We can introduce an \textit{internal} loop within a random walk by extending the procedure outlined in Eq. \eqref{eq:loopedfBm} to an inner segment, which generalize the definition of the bridge function. This enables us to generate sets of fBm-based polymer configurations $\{\vec{x}_n \}$ that incorporate one or more internal loops. These loops are defined by their positions $\iota$ and lengths $\eta$, meaning that monomers $\iota - \eta/2$ and $\iota + \eta/2$ are brought together.

We used these synthetic configurations with internal loops to develop and validate a novel loop-detection technique, known as the $\Lambda$-Plot and based on the computation of the log-spectral ratio.
For each set of $N$-length signals $\{\vec{x}_n\}$, we consider all the sub-signals of length $\eta$, defined as $\{\vec{x}^{(\iota,\eta)}\} = (x_{\iota - \eta/2}\dots x_{\iota + \eta/2-1})$. We calculate the log-spectral ratio $\Lambda(\vec{x}^{(\iota,\eta)})$ for each of these sub-signals and represent the results on a color-scale on the plane $(\iota,\eta)$.
In \Cref{fig:Lambdaplot} we provide typical examples of the expected outcomes when identifying a single inner loop, and compare these results with corresponding distance maps and relevant spectra of sub-polymers.

As showcased in \Cref{fig:Lambdaplot}, $\Lambda$-plots show distinct maxima indicating the presence of a loop. A careful inspection reveals that the $\iota$ coordinate of these maxima precisely corresponds to the midpoint of the loop, while the $\eta$ coordinate is systematically slightly larger than the actual loop size. Thanks to our straightforward loop modeling, we can derive analytical results, as outlined in \Cref{Appendix:maxima}.

For a given fBm signal containing an internal loop centered at $\iota_0$ with a size of $\theta$, the $\Lambda$-plot restricted to the $\iota = \iota_0$ line is indeed given by \Cref{eq:4.13}. This allows a precise determination of the loop position and size starting from the detected maxima $(\iota, \eta)$. As mentioned earlier, we have $\iota_0 = \iota$, and from \Cref{eq:4.13}, the loop size $\theta$ is related to $\eta$ by $\theta = \eta / \mu_0$, where $\mu_0 \approx 1.34767$ is a universal constant. 

With these results, we can formulate a method for detecting loops in signals. Given a set of signals  $\{\vec{x}_n\}$ containing internal loops, to estimate their position and size, follow these steps:
\begin{enumerate}
    \item Calculate the estimated $\Lambda$-plot from the available samples;
    \item Find the position $(\iota=\iota_0, \eta)$ of any maximum;
    \item Divide $\eta$ by $\mu_0 \approx 1.34767$ to find the approximate size of the corresponding loop;
    \item The estimated loop falls then between monomers $\iota_0-\eta/(2\mu_0)$ and $\iota_0+\eta/(2\mu_0)$.
\end{enumerate}
Finally, note that, taken a point $(\eta,\iota)$ on the $\Lambda$-plot, the triangle of which it is the vertex corresponds to the lambda plot of the region $[\eta - \iota/2, \eta+ \iota/2]$, as shown by the multiple examples given in \Cref{fig:Lambdaplot}.


\subsection*{Estimating the ratio of looped to non-looped conformations}

In a typical experimental dataset, only a portion of the configurations will exhibit a specific loop, while the complementary fraction will lack this feature. Consequently, we need to investigate how the log-spectral ratio depends on the probability of occurrence of a given loop, and whether it can provide any information about this probability.
In \Cref{Appendix:mixed} we derive an expression for the log-spectral ratio $\Lambda$ (for fixed $\iota = \iota_0$) for mixed populations in terms of the probability $p$ of having a loop of size $\theta$. From this expression, we learn that the position of the maximum is independent of $p$, while its amplitude depends on it. Since we have acces to this maximal value of the log-spectral ratio from the $\Lambda$-plot, we may use it to know the looping probability $p$. Indeed, inverting this formula yields
\begin{equation}
    p =  \dfrac{\pi^2}{8 \mu_0} \left( 1 - e^{-\Lambda_{\text{max}}} \right) \csc^{2}\left( \dfrac{\pi}{2 \mu_0} \right).
\end{equation}
Crucially, this connection between the fraction of looped conformations and the strength of the maxima provides a means to estimate the fraction of looped conformations in the sample, offering valuable insights for biological datasets.

\subsection*{$\Lambda$-plot loop detection in multiplexed FISH data}

To evaluate the performance of the $\Lambda$-plot method on experimental data, we turned to the MERFISH datasets by \citeauthor{Bintu2018}, acquired from HCT116 cells of human chromosome 21. The examined genomic region spans from 34.6 Mb to 37.1 Mb, sampled at a genomic resolution of approximately 30 kb with a spatial resolution of less than 50 nm (roughly 0.15 kb) \cite{Bintu2018}. Two variants were analyzed: an untreated, wild-type variant, and an auxin-treated variant. Cohesin depletion, resulting from auxin treatment, leads to the removal of TADs at the population level \cite{Rao2014}, without altering the occurrence of TAD-like structures in individual cells. However, it does disrupt the typical positioning of domain boundaries, often associated with CTCF-binding sites, which explains the loss of TADs at the population level \cite{Bintu2018}.

\begin{figure*}[pht]
    \centering
    \hfill
    \begin{subfigure}{.45\textwidth}
      \centering
      \includegraphics[width=.8\textwidth]{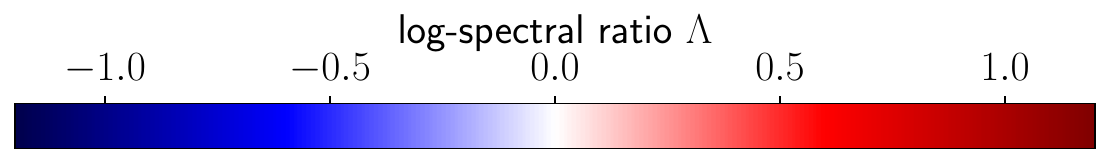}
    \end{subfigure}\hfill%
    \begin{subfigure}{.45\textwidth}
      \centering
      \includegraphics[width=.8\textwidth]{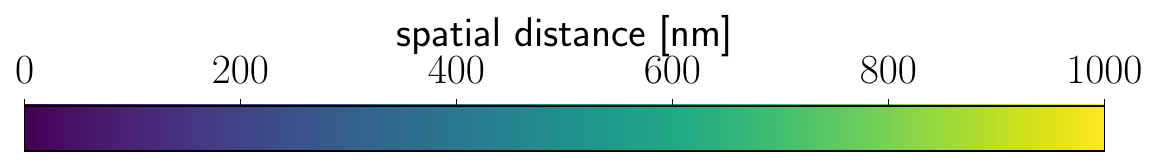}
    \end{subfigure}\hfill
    \begin{subfigure}{.45\textwidth}
      \centering
      \includegraphics[width=.8\textwidth]{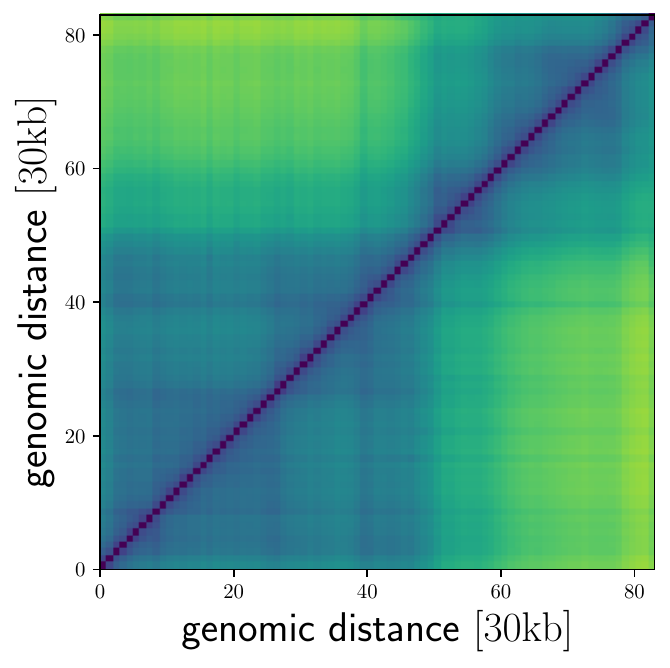}
        \caption{}\label{fig:mapBintu}
    \end{subfigure}%
    \begin{subfigure}{.45\textwidth}
      \centering
      \includegraphics[width=.8\textwidth]{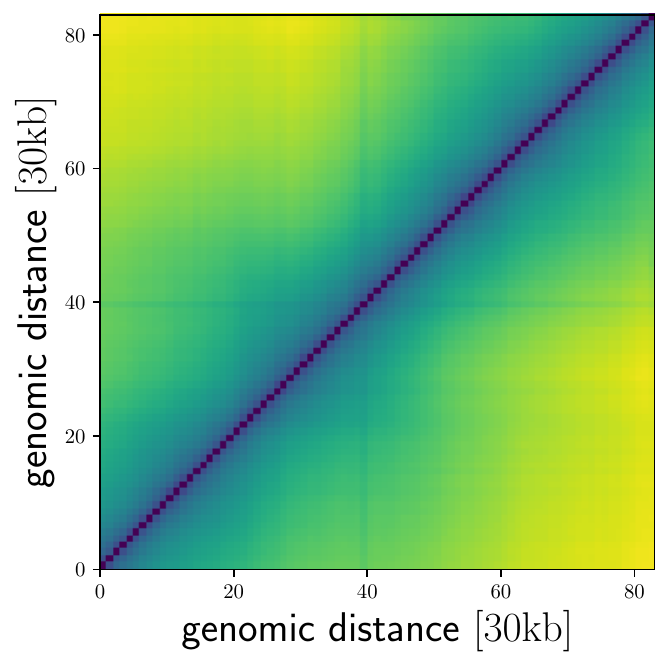}
        \caption{}\label{fig:mapBintu_Auxin}
    \end{subfigure}
    \begin{subfigure}{.45\textwidth}
      \centering
      \includegraphics[width=.8\textwidth]{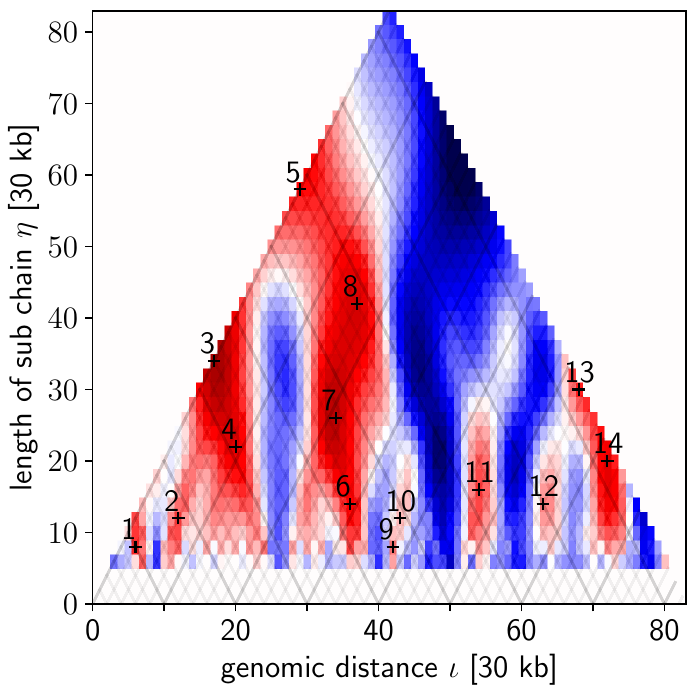}
        \caption{}\label{fig:LambdaBintu}
    \end{subfigure}%
    \begin{subfigure}{.45\textwidth}
      \centering
      \includegraphics[width=.8\textwidth]{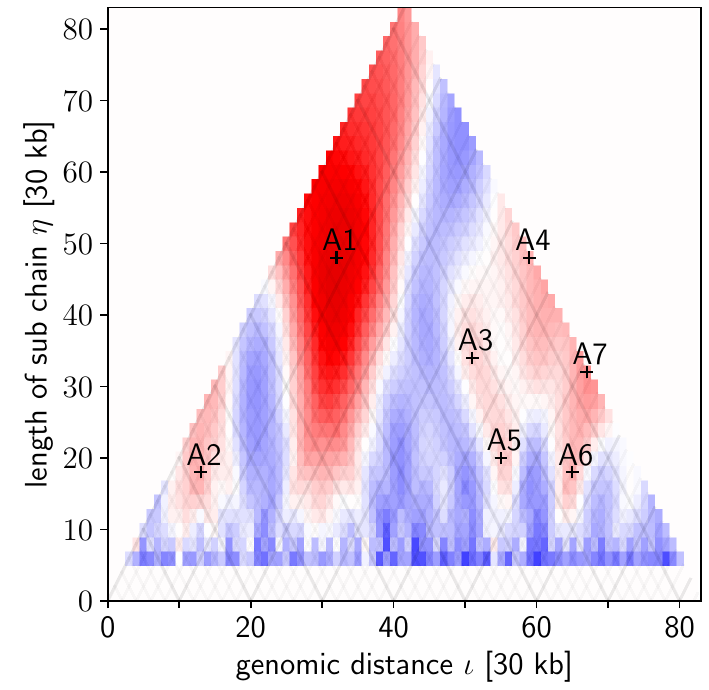}
        \caption{}\label{fig:LambdaBintu_Auxin}
    \end{subfigure}    
    \begin{subfigure}{.45\textwidth}
        \centering
        \includegraphics[width=.8\textwidth]{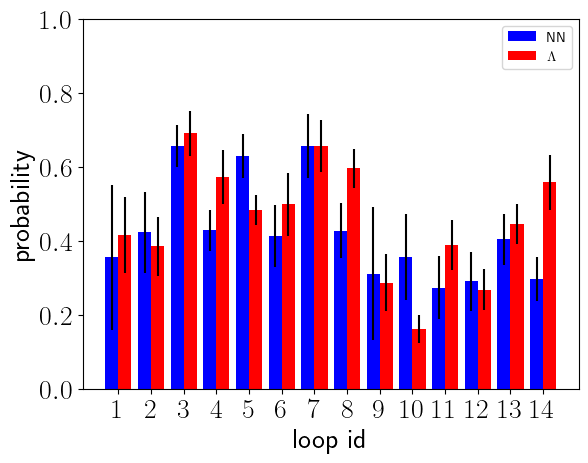}
        \caption{Wild type}\label{fig:histogramsNN1}
    \end{subfigure}%
    \begin{subfigure}{.45\textwidth}
        \centering
        \includegraphics[width=.8\textwidth]{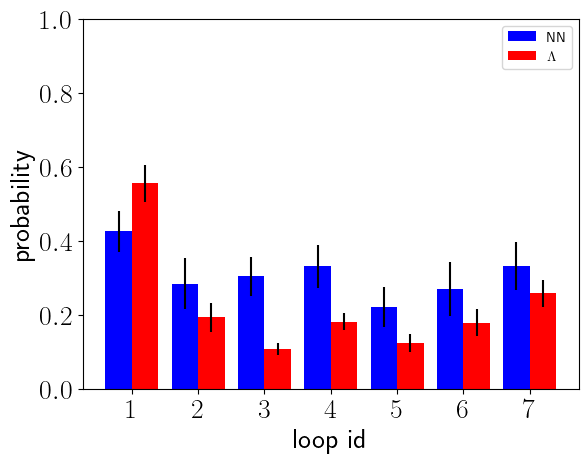}
        \caption{Auxin-treated}\label{fig:histogramsNN2}
    \end{subfigure}
    \caption{The $\Lambda$-plots \textbf{(a,b)} and distance maps \textbf{(c,d)} created from the experimental data of \citeauthor{Bintu2018} for both the wild type variant (left) and an Auxin-treated variant (right). Each detected maximum is given a loop id. In \textbf{(e,f)}, the estimated probability of each loop's occurrence is shown, obtained from the $\Lambda$-plot (red bars) and the neural network output (blue bars). }
\end{figure*}

In \Cref{fig:mapBintu,fig:mapBintu_Auxin}, we first present 
the average distance maps we obtained for the two data sets. In the wild-type data, two large TADs are evident, along with numerous sub-TADs. However, identifying specific loops is challenging. 
In the auxin-treated variant, the (sub-)TADs are less pronounced, and a significant loss of structural detail is observed at the ensemble average level. No distinct loop can be identified from the distance map.

The $\Lambda$-plots for the same data sets, along with the identified maxima and loop sizes, are presented in \Cref{fig:LambdaBintu,fig:LambdaBintu_Auxin}. 
The log-spectral ratio successfully detects numerous loops in the conformations, including 14 loops in the wild type (labeled 1 to 14) and 7 loops in the auxin-treated variant (labeled A1 to A7). Maxima detection involves manual selection of regions where they might be present, followed by standard numerical methods. The estimated proportions of looped conformations for each loop, along with their respective errors, are summarized in \Cref{fig:histogramsNN1,fig:histogramsNN2} (red bars). 

\begin{table*}[tbh]
\centering
\begin{tabular}{c|ccccccc}
loop id    & 1 & 2 & 3 & 4 & 5 & 6 & 7 \\ \hline
$\iota_0$  & 6 & 12 & 17 & 20 & 29 & 36 & 34 \\
$\eta$        & 8 & 12 & 34 & 22 & 58 & 14 & 26 \\
Loop range & \phantom{0}3-9\phantom{0} & \phantom{0}8-16 & \phantom{0}4-30 & 12-28 & \phantom{0}7-51 & 31-41 & 24-44
\end{tabular}\vspace{2mm}\\
\begin{tabular}{c|ccccccc}
loop id    & 8 & 9 & 10 & 11 & 12 & 13 & 14 \\ \hline
$\iota_0$  & 37 & 42 & 43 & 54 & 63 & 68 & 72 \\
$\eta$        & 42 & 8 & 12 & 16 & 14 & 30 & 20 \\
Loop range & 21-53 & 39-45 & 39-47 & 48-60 & 58-68 & 57-79 & 65-79
\end{tabular}\vspace{2mm}\\
\begin{tabular}{c|ccccccc}
loop id  & A1 & A2 & A3 & A4 & A5 & A6 & A7\\ \hline
$\iota_0$  & 32 & 13 & 51 & 59 & 55 & 65 & 67 \\
$\eta$        & 48 & 18 & 34 & 48 & 20 & 18 & 32 \\
Loop range & 14-50 & \phantom{0}6-20 & 38-64 & 41-77 & 48-62 & 58-72 & 55-79 
\end{tabular}
\caption{ The $\iota_0$ and $\eta$ coordinate of the maximum in the $\Lambda$-plots for each loop identified in \Cref{fig:LambdaBintu}. The inferred loop extremities $\iota_0 \pm \eta / (2 \mu_0)$ are also listed. } 
\end{table*}

Examining the relative positions of loops is also interesting. Some loops overlap or are included within larger loops,  as can be understood by visualizing the corresponding triangles in the $Lambda$ diagrams.
For example, loops 12 and 13 are inside loop 14, while loop 11 is relatively isolated from the others. In the auxin case, loops A3, A5, A6, and A7 are within loop A4, and loop A1 partially overlaps with loop A2.

Notably, some loops are closely adjacent to each other, such as loops A3 and A7 or loops 12 and 14, forming what appears to be the two "petals" of a flower-like shape. It's interesting to note that Ref. \cite{Espinola2021} suggests that flower-like looping is a fundamental mechanism in chromatin folding leading to hubs or
cluster of interacting cis-regulatory modules including enhancers and promoters.
This suggests that our algorithm is capable of detecting such structures. However, it's important to confirm that these loops are present simultaneously in unique configurations, rather than being a result of averaging across the entire dataset.
To address this question, we need to determine in which specific samples a detected loop is present. This will be explored in the next subsection.

\subsection*{Using neural networks to segregate looped and non-looped configurations}

\begin{figure*}[t]
    \centering
    \includegraphics[width=\textwidth]{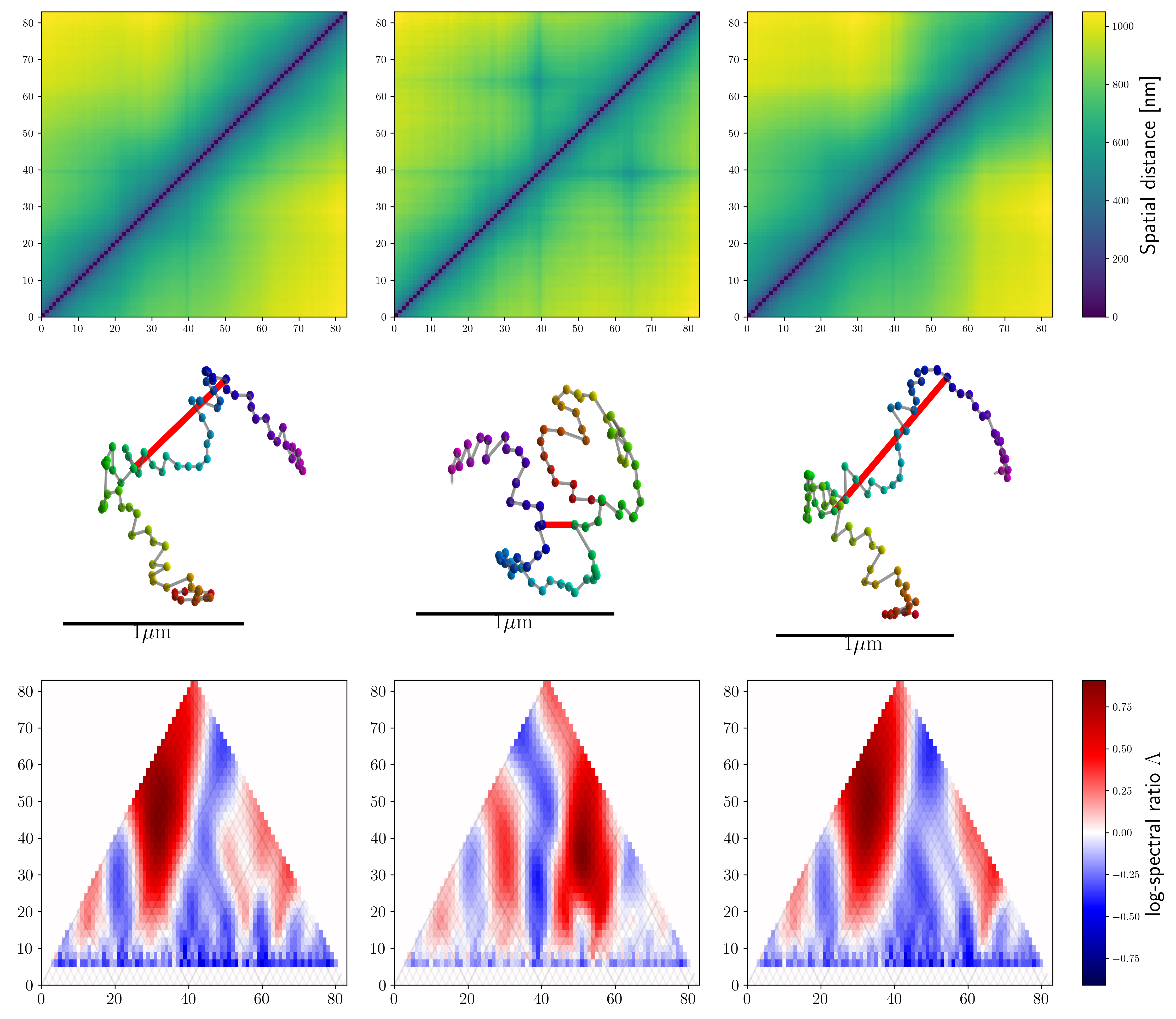}
    \caption{Typical example of output of neural network segregation of looped and non-looped populations, for loop A3. The first column shows the distance map, $\Lambda$-plot and mean configuration (similar to the ShRec3D algorithm \cite{Lesne2014}, see \nameref{sec:methods}) for all measurement data. The second and third columns show distance map, $\Lambda$-plot and mean configuration for measurements that the NN recognized as containing or lacking loop A3, respectively.  }
    \label{fig:segregation}
\end{figure*}

To validate and complement our log-spectral method, we developed a neural network (NN) approach to assess the presence of specific loops in individual conformations. To this aim, we introduced a neural network (NN) approach. Our neural network was trained using artificially generated looped and non-looped random walks, employing 20,000 training samples, 5,000 validation samples, and 2,000 test samples in each category. Importantly, generating our own training data is highly advantageous, as it avoids using experimental data for network training, effectively minimizing data wastage. Overfitting is controlled, and the test samples provide a reliable estimate of the neural network's accuracy. Comprehensive details are provided in \Cref{Appendix:NN}.

To independently gauge the presence of loops, our NN is fed with the spatial distances between pairs of points equidistant from the loop midpoint $\iota$ along the chain. In looped configurations these distances should exhibit a minimum at approximately half the loop's length, while in non-looped ones, on average, they should show a linear increase with distance (see \Cref{fig:NN_input}).
Once a loop is identified, by locating a maximum $(\iota, \eta)$ in the $\Lambda$-plot of FISH data, we use the trained NN on each individual conformation to ascertain whether it contains a loop at the specified position. For details, see \Cref{Appendix:NN}.
The neural network approach offers the added benefit of allowing us to collect data at the single-cell level, enabling further analyses on segregated datasets. We emphasize that the $\Lambda$-plot is pivotal for the NN's applicability, since the NN can only be applied to one location at a time and is specifically trained for a single loop size.

As a first test,  we compare the proportions of looped conformations determined by the neural networks to those obtained from the $\Lambda$-plot for each detected loop in the data sets. The results are presented in \Cref{fig:histogramsNN1,fig:histogramsNN2}. Hypothesis testing reveals that we can reject the null hypothesis of equal estimated proportions for all loops except for loops 14 and A3 (or 5, 14, A3, and A4) with 99\% (or 95\%) confidence. This strong agreement between the two methods underlines their reliability.

The NN approach enables precise discrimination at the individual conformation level once a loop is detected in the population by the lambda-plot method. This, in turn, enables the separation of two distinct sub-populations: one with looped configurations and the other with non-looped configurations. 
For illustration, in \Cref{fig:segregation}, we present a comparison of average distance maps for the whole Auxin-treated dataset and those derived from its sub-populations - one with loop A3 ($\iota_0=51,\eta=34$) and the other without, as determined by our NN approach. 

Strikingly, in the looped sub-population distance map, a local minimum, a typical indicator of loops in contact maps, appears at the position of the predicted loop. 
Correspondingly, the $\Lambda$-plots show a very strong enhancement of the A3 maximum in the looped population, where it overcomes all other maxima, while it is clearly suppressed in the plot for the non-looped population.

Additionally, we include the corresponding mean configurations, reconstructed following the method outlined in \nameref{sec:methods}, which provides additional confirmation of the NN's effectiveness in distinguishing configurations containing a loop within the region pinpointed by the $\Lambda$-plot approach. 

To further validate the method's accuracy, we implemented the NN procedure on regions identified by the $\Lambda$-plots as lacking loops. The results are discussed in \nameref{sec:methods} and demonstrate the NN's capability to correctly discern the absence of a significant looped sub-population.

\subsection*{Anti-correlation between adjacent loops detected}
\begin{figure*}[tb]
    \centering
    \begin{subfigure}{.48\textwidth}
        \centering
        \includegraphics[width = \linewidth]{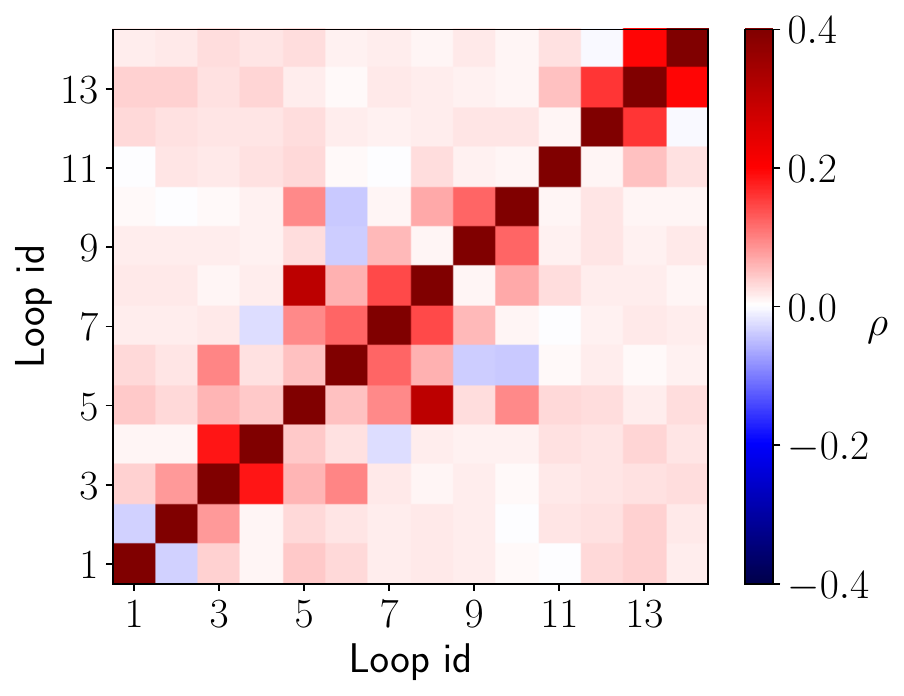}
        \caption{Wild type}\label{fig:correlations1}
    \end{subfigure}%
    \begin{subfigure}{.48\textwidth}
        \centering
        \includegraphics[width = \linewidth]{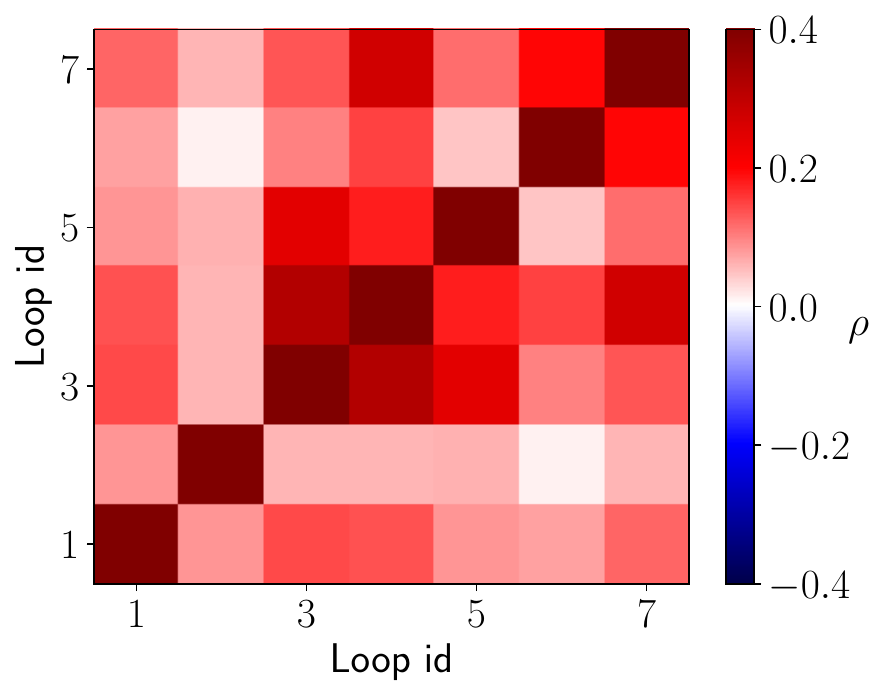}
        \caption{Auxin}\label{fig:correlations2}
    \end{subfigure}
    \begin{subfigure}{.67\textwidth}
        \centering
        \includegraphics[width = \linewidth]{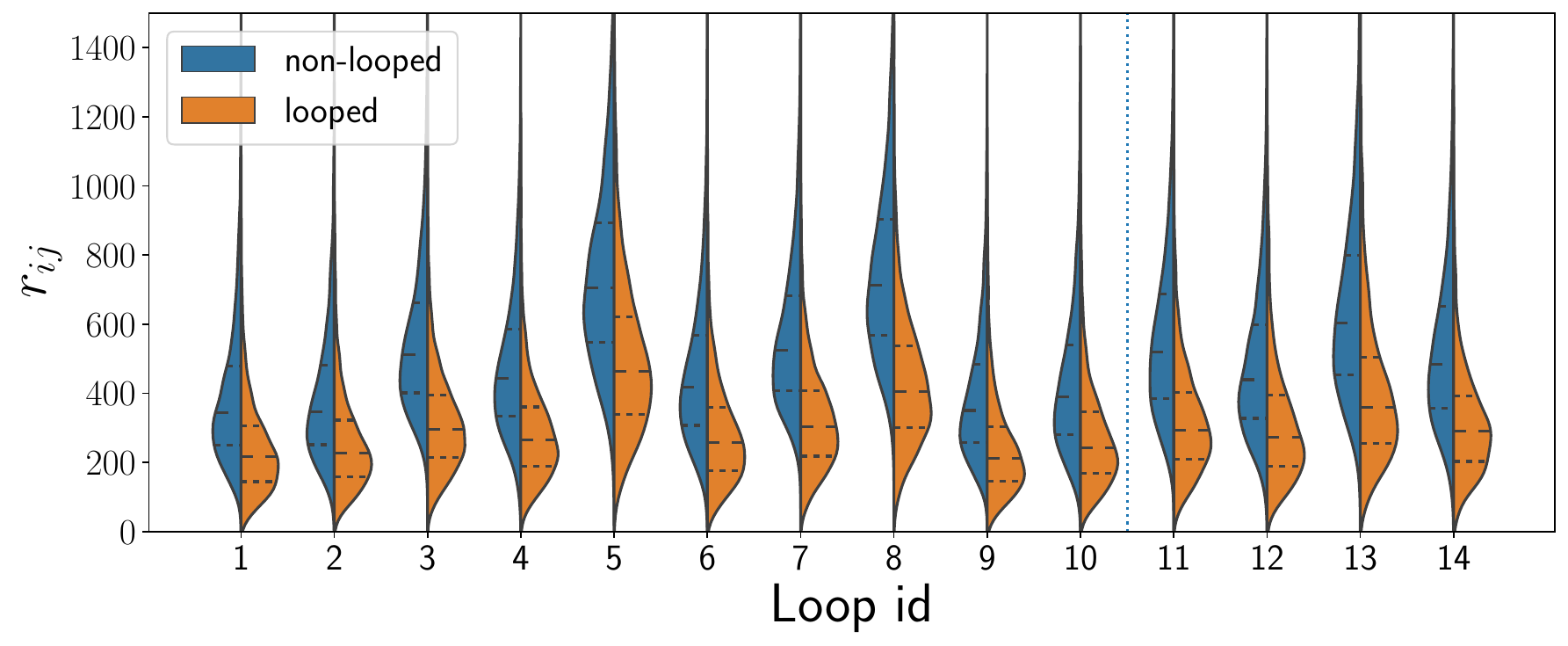}
        \caption{Wild type}\label{fig:violinplots_distances1}
    \end{subfigure}%
    \begin{subfigure}{.33\textwidth}
        \centering
        \includegraphics[width = \linewidth]{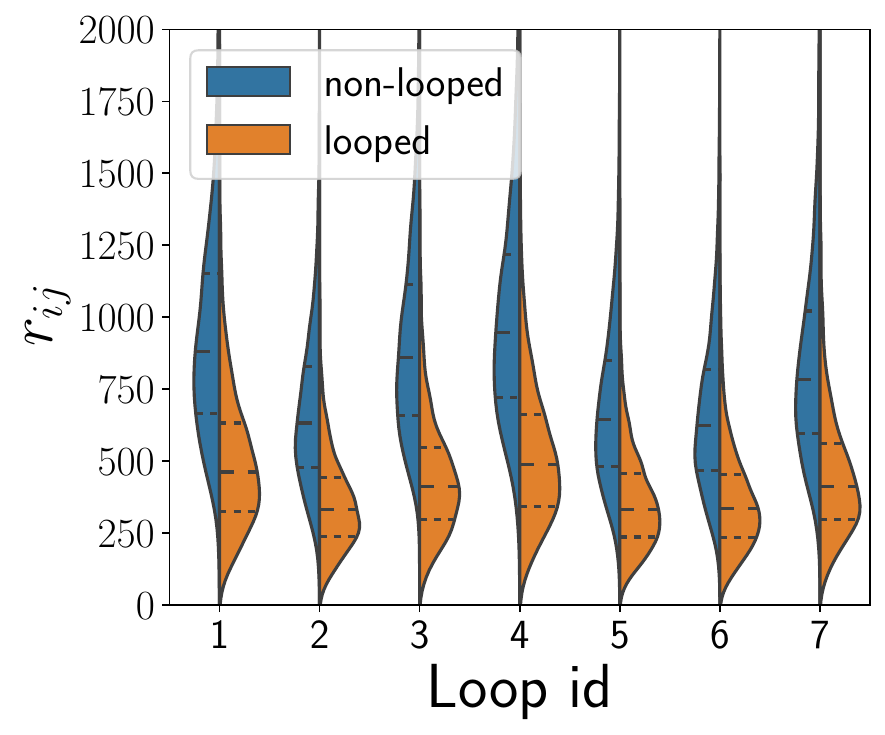}
        \caption{Auxin}\label{fig:violinplots_distances2}
    \end{subfigure}
    \caption{\textbf{(a,b)} Pearson correlation between loops in both wild type and Auxin-treated data sets.
    \textbf{(c,d)} Distributions of end-to-end distances $r_{ij}$ which measure the separation between the two extremities $i=\iota_0-\eta/(2\mu_0)$ and $j=\iota_0+\eta/(2\mu_0)$ for looped (orange) and non-looped (blue) configurations across all loops identified in the FISH data.}
\end{figure*}

By separating the looped and non-looped conformations, we've been able to investigate the relationships between loops, specifically the joint probabilities of each loop pair. In \Cref{fig:correlations1,fig:correlations2}, we present the Pearson correlations for the loops in the experimental data.
All loops in Auxin-treated, except loop A2, are positively correlated with each other. However, loops A3, A4 and A7 seem to be correlated to each other pairwise, 
consistent with the idea of A3 and A7 forming the two petals of a flower like shape, where the combination of A3 and A7 is the loop A4.
Similarly, on the wild type variant, loop 13 is the combination of loops 12 and 14. Loops 12 and 13, as well as 13 and 14, are positively correlated, while loops 12 and 14 are anti-correlated. This suggests that the flower-like shape is less likely to occur than the two individual loops separately. Instead, it seems the flower-like shape only emerges from averaging over multiple cells.

In the wild type, it is also remarkable that loop 11 seems to be independent of the other loops. 
Furthermore, it's interesting to observe anti-correlations between neighboring loops, such as loops 1 and 2, loops 6 and 10, and loops 12 and 14. This might suggest an underlying biological mechanism that prevents adjacent loops from occurring simultaneously.

\subsection*{End-to-end distance distributions differ for looped and non-looped populations}

The distributions of end-to-end distances - distance between the two extremities of the loop - are shown in \Cref{fig:violinplots_distances1,fig:violinplots_distances2}, and reveal variations between looped and non-looped populations in the FISH data for both the wild type and Auxin-treated cases. These populations are considered separately for comparison. In the looped population, a prominent peak at shorter distances is evident, whereas the non-looped population exhibits a broader distribution centered on larger distances and growing with the loop size, in agreement with what expected for linear polymers.

It's important to emphasize that there is a significant overlap in the end-to-end distance distributions between these two populations. This finding demonstrates the inadequacy of a simple analysis of inter-loci distances for loop discrimination and point to the need for a more comprehensive approach, as proposed in this study.

\subsection*{Further insights into chromatin architecture} 

We can use the analogy of fBm to gain further insights into chromatin architecture features in TADs.
Let's consider the two large TADs in the wild-type (from 0 to  $50\cdot 30 \text{kb}$, region (1), and $50\cdot 30 \text{kb}$ until  $83\cdot 30 \text{kb}$, region (2))  and the entire region in the Auxin-treated dataset (Region 3) as a potential third TAD.
If we treat these regions as non-looped, we can fit the internal end-to-end distance $R(s)$ with a power law $f(s) = A (s/30  \text{kb})^H$, for each of these regions. 
The fitted exponents $H$ are given in the first row of \Cref{tab:Hfitted}. 
It's worth noting that these three values are quite close to each other, and their exponents are not significantly distant from 1/3, which is the typical exponent expected for the crumpled globule model \cite{Grosberg1988,Mirny2011}.

However, our previous results allows us to potentially determine the effects of the presence of loops on the exponent $H$. 
In particular, if we only select the population with two loops or fewer for the wild-type, and the population without any loops for the Auxin-treated variant, we find different exponents, as shown in the second row of \Cref{tab:Hfitted}.
By excluding looped populations, the fitted exponents change notably, becoming closer to 0.4 rather than 0.3. This suggests that an incorrect interpretation of $R(s)$ behavior in experimental data might result from the influence of undetected loops in the chromatin.

It's important to note that an exponent of about 0.4 in non-looped chromatin is consistent with the hypothesis that chromatin conformations can be described as polymers at the coil-globule phase transition \cite{Lesage2019, Foldes2021}, which indeed leads to a wider range of possible exponents.
In any case, it is worth noting that our approach  allows us to segregate looped and non-looped populations, enabling a more accurate interpretation of the data.

\begin{table*}[tbh]
    \centering
    \begin{tabular}{|c|c|c|c|} \hline 
         region &  (1)&  (2)& (3)\\ \hline 
         all conformations&   $0.283 \pm 0.014$&  $  0.314 \pm 0.014$&  $ 0.300 \pm  0.003$\\ \hline 
         low loop content&  $0.394 \pm  0.015$&  $ 0.418 \pm  0.016$& $0.411 \pm  0.005$\\ \hline
    \end{tabular}\vspace{2mm}
    \caption{Values of the exponent $H$ obtained by fitting $R(s) = A (s/30 kb)^H$ for regions (1), (2) and (3) (see main text) while considering all the conformations (\textbf{upper row}) or only conformations with two loops or less (for the wild type) or without any loops (for the Auxin-treated variant) (\textbf{lower row}).}
    \label{tab:Hfitted}
\end{table*}

\section*{Discussion}\label{sec:discussion}

The conventional methods of loop detection via distance maps failed to notice the presence of multiple loops in the experimental multiplexed super-resolution FISH data of \cite{Bintu2018}.
Furthermore, FISH experiment data offers a more comprehensive information than distance maps by encompassing the complete 3D configuration, information that has been overlooked until now.
The $\Lambda$-plot proposed in this work accomplishes this task.
By using the whole amount of information that is available from these FISH experiments and not just the average distances between the markers, the $\Lambda$-plot approach clearly indicates the presence of these loops in the ensemble of measured chromatin configurations. It provides a reliable and fast method to detect loops in FISH data, irrelevant of size and position, and sensitive to small looped populations.

The presence of loops is confirmed via a neural network approach, which further results in the opportunity to classify chromatin as looped or non-looped in each cell. This classification is achieved by assessing the presence of specific loops in each measurement. We have demonstrated the feasibility, speed, and reliability of this process. A significant portion of the success of this neural network approach is attributed to the initial guidance provided by the $\Lambda$-plot and the ease with which we can generate artificial training data based on an fBm model. As a result, we can avoid wasting valuable measurements.

The method introduced here broadens data processing possibilities and strengthens the foundation for advancing chromatin's theoretical understanding through precise and comprehensive experimental data analysis. Our analysis, for instance, prompts a critical reevaluation of the crumpled globule model. We discovered that the corresponding critical exponents of $1/3$, frequently encountered in experimental data, may result from averaging looped and non-looped configurations within a dataset.
This effect may have remained unnoticed, due to the lack of an efficient loop detection procedure. 

Another intriguing finding involves the potential existence of clusters of adjacent loops, resulting in flower-like structures reminiscent of cis-regulatory module hubs \cite{Espinola2021}. The ability to examine loops at the single-cell level now allows for a quantitative investigation of correlations between different loops for the first time. Our results suggest that the presumed occurrence of neighboring loops forming a flower is also the result of an average between looped and non-looped conformations, in the investigated dataset.

Multiple future research endeavors are possible with the developed method, and we highlight some topics that warrant further exploration. First of all, the biological function of the detected loops requires further research. In particular, the abundance of loops in the Auxin-treated variant raises questions about their relevance and potential roles. The method we developed represents a necessary initial step to study these loops. 
Secondly, it is certainly necessary to study in detail which proteins are present at the boundaries of the detected loops. Firstly, the correlation with CTCF and cohesin needs to be examined. A more detailed study may identify new proteins responsible for loop maintenance.
Thirdly, a technical aspect of the method can be enhanced: by Employing pattern recognition techniques could improve the detection of peaks and associated loops within the $\Lambda$-plot. 
Fourthly, our confidence in the versatility of the spectral-based technique developed in this study encourages its application to investigate a broader range of phenomena.  For instance, the method can be adapted for detecting plectonemes in supercoiled DNA or for identifying density variations across the genome or in spatial arrangements, such as alternating coils and globules, or alternating A and B compartments. These structures are predominantly characterized by their large-scale behaviors, where low-mode spectral features prove to be particularly suitable for in-depth investigation.
Preliminary investigations of data from Ref. \cite{Su2020}, that is 
 on a much larger scale than the one considered here, seems to indicate indeed that the same analysis can readily identify AB-compartments and their corresponding boundaries. These findings are consistent with the conclusions drawn in the original paper. Additionally, loops were also detected and warrant further investigation in future research.

\section*{Methods}\label{sec:methods}

\begin{figure*}[!tb]
	\centering
    \begin{subfigure}{.5\linewidth}
        \centering
    	\includegraphics[width = .75\linewidth]{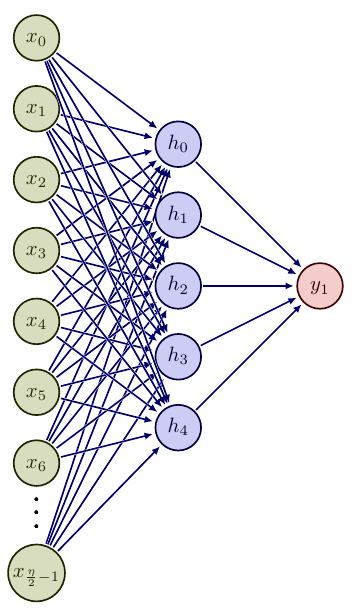}
        \caption{}\label{fig:NN_separating_looped_non_looped_polymers}
    \end{subfigure}%
    \begin{subfigure}{.5\linewidth}%
        \centering
        \includegraphics[width=.75\linewidth]{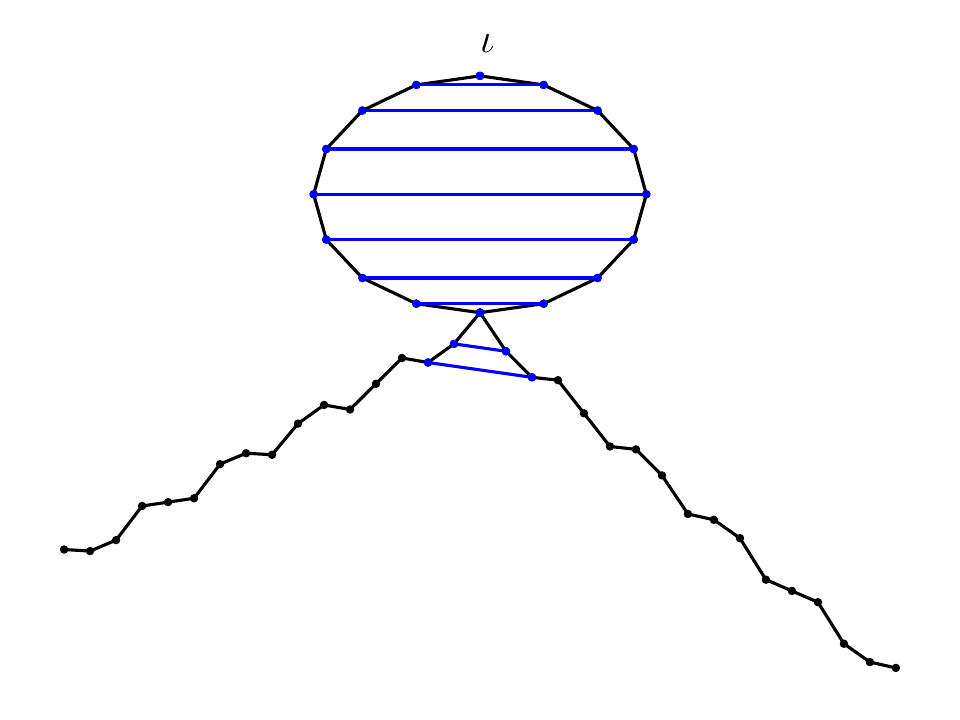}
         	\includegraphics[width=\linewidth]{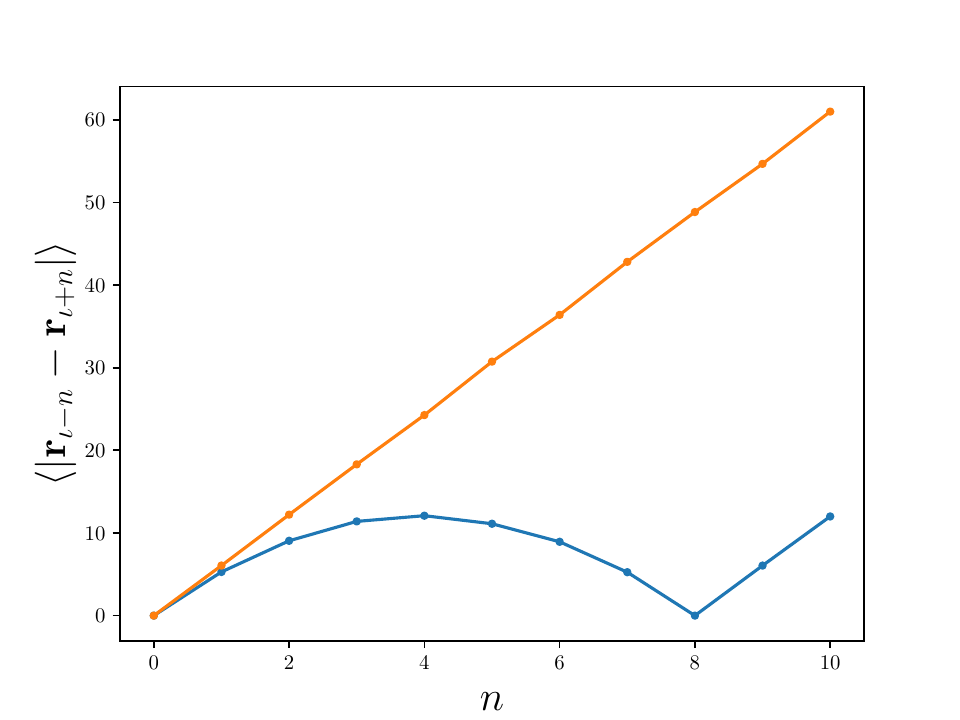}
         \caption{}\label{fig:NN_input}
         \end{subfigure}
    	\caption{ \textbf{(a)} The neural network we build to separate looped and non-looped polymers. We input the information of the polymer as sketched in (b top), and the output is a value between 0 and 1 representing the probability that the polymer that was put in is a looped one. The activation function of the hidden layer is the ReLu function, and the sigmoid activation is used for the output layer. 
        \textbf{(b top)} As input of the neural networks, we choose to use the spatial distances between points that are an equal chain distance away from the loop midpoint $\iota$. \textbf{(b bottom)} We show the expected input for looped (blue) and non-looped (orange) random walks, obtained from averaging over 3000 random walks and 3000 looped random walks. }
\end{figure*}

\subsection*{Neural networks specifics}\label{appendix:NN_specifics} 

Each time the position of a maximum $(\iota, \eta)$ of the $\Lambda$-plot is found, a neural network as in \Cref{fig:NN_separating_looped_non_looped_polymers} is trained to separate random walks of length $\eta$ with and without internal loop. The loop sizes lie uniformly in the range \[ \frac{\eta}{\mu_0} \pm \max\left( 0.1 \frac{\eta}{\mu_0}, 1 \right), \] where $\mu_0$ is given in \Cref{eq:value_mu_0}. This range is arbitrarily chosen as to give enough variability in the training data so that the neural network can more easily generalize to unseen data. The network has the 
ReLU-activation function on the hidden layer, and the sigmoid-activation function on the output-layer, see for example \cite{aggarwal_neural_2018}. We use the binary cross-entropy as the loss function.

The polymer data is inputted as follows. Since the midpoint of the loop is the most certain prediction of the $\Lambda$-plot, the distances between the markers and this midpoint are studied. In formulae, we want to represent the (looped) random walk $\{\vec{r}_i\}$ as the signal $\vec{\widetilde{x}}$ given by
\[ \widetilde{x}_n = \abs{ \vec{r}_{\iota - n} - \vec{r}_{\iota + n} } \textrm{ for } n = 0, \hdots, \eta/2-1.  \] See \Cref{fig:NN_input} top for a sketch of the situation and \Cref{fig:NN_input} bottom for expected outputs of $\vec{\widetilde{x}}$ for both looped and non-looped random walks. Since the signal $\vec{\widetilde{x}}$ still has an inherent scale to it, we normalize it to obtain $\vec{x}$ as 
\[ x_n = \frac{\widetilde{x}_n}{\max(\vec{\widetilde{x}})}. \] This way, we try to lessen the effect of compact versus loose packing of the chromatin, as well as that of small and large loops. 

We train our neural network on artificially generated looped and non-looped random walks---with balanced training data---with 20000 learning samples, 5000 validation samples, and 2000 test samples for both the looped and non-looped random walks. The validation samples are used to monitor and prevent overfitting, and the test samples give an estimate of the accuracy of the neural network. It needs to be remarked that it is an enormous benefit that we can artificially generate training data, as we do not need to waste any experimental data on training the networks.

\begin{figure*}
    \centering
\includegraphics[width=.5\linewidth]{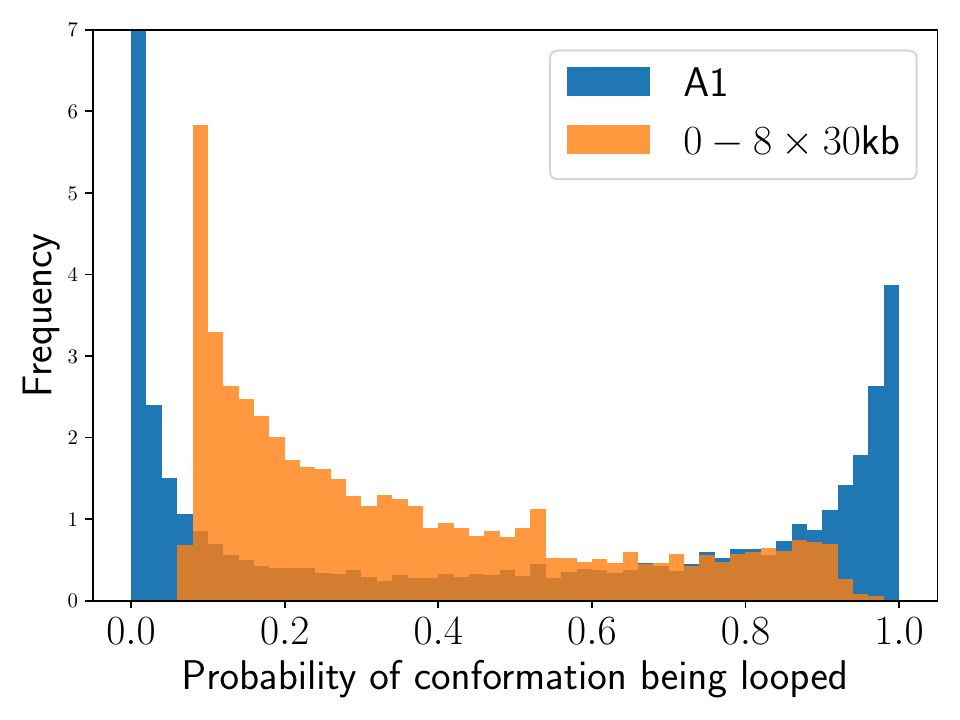}
    \caption{Histogram of probability of a single measurement configuration being looped, for loop region A1 and for the non-looped region $0-8 \times 30\text{kb}$. }
    \label{fig:segregation_wrong_region}
\end{figure*}

\subsection*{Mean polymer configuration: ShRec3D-like approach}\label{Appendix:SHREK}

The ShRec3D algorithm \cite{Lesne2014} is aimed to reconstruct spatial distances and three-dimensional genome structures from observed contacts between genomic loci. In the data from multiplexed super-resolution, the single configurations are known. However, we follow a simplified approach in the spirit of the ShRec3D algorithm in order to have a representation of the \textit{average} features of an entire data set. 
To this aim, we calculate individual distance maps for each configuration, then average over all these maps. This average map will be invariant to translations and rotations of each individual polymer. Moreover, the averaged map will still be a distance map (i.e. be symmetric and satisfy the triangle inequalities). Hence, we can choose to put the first monomer in the origin, the second monomer on the positive $x$-axis, and the third monomer on the $z=0$-plane, and then the distance map completely determines the polymer configuration. This is then the average configuration.

\subsection*{Neural network applied to non-looped regions}\label{Appendix:non_looped_NN_application} 

To check the occurrence of false positives in the NN loop detection,  we select a random region (from $0\text{kb}$ until $8 \times 30 \text{kb}$) which, according to the $\Lambda$-plot, contains no loops. Note moreover that this is a small region, which is generally more difficult for the neural network to work with. \Cref{fig:segregation_wrong_region} shows the estimated loop-probability (output of the NN) for the selected region and, for comparison, for the region of loop A1. The two outputs are qualitatively different. The distribution of the looped region is bimodal, indicating the presence of a looped sub-population  while that of the random region has a single modus. Moreover, the random region displays a steep cut-off before reaching a probability of one of being looped.

\subsection*{Time complexity}\label{appendix:time_complexity} 

Creating the $\Lambda$-plot requires studying all the sub-polymers at all possible positions, which can be quite time extensive at first glance. Luckily, due to application of the Fast Fourier Transform to compute the Discrete Cosine Transform, and by using the fast vectorizing abilities of numerical software like NumPy, this is actually not a problem. Without performing a detailed analysis -- since the timing results were satisfactory -- we can report that the creation of the two experimental $\Lambda$-plots of \Cref{fig:LambdaBintu} only took about 30 seconds, which is for around 20000 configurations of 83 3D-points each. This timing is for a MacBook pro with apple M1 MAX chip and 32 GB RAM. The training and application of each neural network to each separate loop takes about 11 minutes in total (running in parallel with 10 cores).


\bibliography{Liefsoens-bibliography}

\newpage
\onecolumn
\begin{appendices}

\crefalias{section}{appsec}

\section{Spectrum of looped correlated random walks}\label{Appendix:loopedPSD}

Starting from a fBm signal $\boldsymbol{\gamma}_n$, a looped fBm is defined as \cite{Gasbarra2007}  
\begin{equation}
\label{eq:loopedfBm_app}
    \boldsymbol{\lambda}_n =\boldsymbol{\gamma}_n - {\cal B}_H(n;0,N)\, \vec{R}
\end{equation}
where $\vec{R}=\boldsymbol{\gamma}_N-\boldsymbol{\gamma}_1$ is the fBm end-to-end vector and  
\begin{equation}\label{eq:bridge_function_defined}
   {\cal B}_H(n;0,N) 
={N^{-2H}}\, C_{\gamma\gamma}(n,N)
\end{equation}
is the appropriate bridge function needed to connect the two ends of fBm to construct an fBm loop. 
Given this model of a looped random walk, we can calculate the corresponding PSD.

We will write $\left\langle \widehat{\mathcal{E}}_p(\vec{\lambda}) \right\rangle$ for the PSD of $\vec{\lambda}$, so this is the expectation value of the square value of the DCT of $\vec{\lambda}$. Analogously, we will denote the PSD of $\vec{\gamma}$ by  $\left\langle \widehat{\mathcal{E}}_p(\vec{\gamma}) \right\rangle$. 

The linearity of the DCT gives that the PSD of $\vec{\lambda}$ is given as 
\begin{equation}\label{eq:in_der_1_looped_corr_RW_vs_RW}
    \dfrac{\left\langle \widehat{\mathcal{E}}_p(\vec{\lambda}) \right\rangle}{
    \left\langle \widehat{\mathcal{E}}_p(\vec{\gamma}) \right\rangle} = 1 - \frac{1}{\left\langle \widehat{\mathcal{E}}_p(\vec{\gamma}) \right\rangle} \left(\DCT_{ \left( \mathcal{B}_h(n; 0, N) \right)_{n=1,\hdots,N} }(p)\right)^2 \langle \vec{R}^2 \rangle,
\end{equation}
making explicit the relation between the spectra of the correlated random walk $\vec{\gamma}$ and the looped variant $\vec{\lambda}$.  Note that it is simply the bridge function that determines the difference in PSD between the looped and non-looped fBm. 

For the DCT of the bridge function $\left( \mathcal{B}_h(n; 0, N) \right)_{n=1,\hdots,N}$ of \Cref{eq:bridge_function_defined}, we have per definition
\begin{equation}
    \DCT_{ \left( \mathcal{B}_h(n; 0, N) \right)_{n=1,\hdots,N} }(p)
    =  \frac{1}{2N^{2H+1}} \sum_{n=1}^{N} \left( n^{2H} - (N-n)^{2H} \right) \cos\left( \frac{p \pi}{2 N}(2n-1) \right)
\end{equation}
for $p>0$. We cannot further simplify this summation, as the exponent $2H$ is a real number, and no longer an integer. Therefore, we rewrite it as
\begin{align*}
    \DCT&_{ \left( \mathcal{B}_h(n; 0, N) \right)_{n=1,\hdots,N} }(p) \\
    &=  \frac{1}{2} \sum_{n=1}^{N} \frac{1}{N} \left( \left(\frac{n}{N}\right)^{2H} - \left(1-\frac{n}{N}\right)^{2H} \right) \cos\left( p \pi\left( \frac{n}{N}-\frac{1}{2N}\right) \right),
\end{align*}
and convert this summation to an integral, valid for $N \gg 1$:
\begin{equation}\label{eq:in_der_2_looped_corr_RW_vs_RW}
    \DCT_{ \left( \mathcal{B}_h(n; 0, N) \right)_{n=1,\hdots,N} }(p)
    \approx  \frac{1}{2}  \int_0^1 \left( x^{2H} - (1-x)^{2H} \right) \cos\left( p \pi x \right) \,\dd x.
\end{equation}
Note that $1/N$ of the summation became the volume element in this integral. We will show that this integral is zero for even $p$, so that \Cref{eq:in_der_1_looped_corr_RW_vs_RW} will give that the looped correlated random walk has the same even modes as the ordinary correlated random walk, at least for long chains.

Denote the integrand of \Cref{eq:in_der_2_looped_corr_RW_vs_RW} by the function $g_p$:
\[ g_p(x) = \left( x^{2H} - (1-x)^{2H} \right) \cos\left( p \pi x \right) \]
and note its symmetry:
\begin{equation*}
    g_p(1-x) = (-1)^{p+1} g_p(x).
\end{equation*}
From this symmetry, the integral of \Cref{eq:in_der_2_looped_corr_RW_vs_RW} becomes
\begin{equation*}
    \int_0^1 g_p(x) \,\dd x 
    = \int_0^1 g_p(1-y) \,\dd y
    = (-1)^{p+1} \, \int_0^1 g_p(y) \,\dd y,
\end{equation*} after a change of variables $y=1-x$.
From this, we can conclude 
\begin{equation*}
    \left( 1 + (-1)^p\right) \int_0^1 g_p(x) \,\dd x = 0 \\
\end{equation*}
or
\[ \int_0^1 g_p(x) \,\dd x = 0 \quad \text{for } p \text{ even}. \]
From this last equation, we can conclude from \Cref{eq:in_der_1_looped_corr_RW_vs_RW} that
\begin{equation}
    \left\langle \widehat{\mathcal{E}}_p(\vec{\lambda}) \right\rangle = 
    \left\langle \widehat{\mathcal{E}}_p(\vec{\gamma}) \right\rangle \qquad \text{for } p \text{ even},
\end{equation}
i.e. the even modes of looped and non-looped (infinite length) correlated random walks are the same. This is a direct consequence of the symmetry of the bridge function of \Cref{eq:bridge_function_defined}.

Let us now consider a general ideal circular signal $\vec{x}$. Then, the first point $x_0$ of the signal $\vec{x}$ is equal to the last point $x_{N-1}$. By the symmetry of the DCT operation, it follows that 
\begin{equation}\label{eq:constraint_odd_modes_RW_loops}
    \sum_{\substack{p=1\\ p \text{ odd}}}^{N-1} X_p \cos \left( \frac{p\pi}{2N} \right) =0.
\end{equation}

This constraint states that the weighted sum of the odd modes should be zero, and is of topological nature. So, if the first mode is large, the other odd modes have to compensate for this by being small. On average this leads to the lowering of all the odd modes. Since $\cos \left( \frac{p\pi}{2N} \right)$ is a decreasing function in $p$, the first mode has the most effect on satisfying this constraint, and goes down the most, relatively speaking.

\section{Definition of \texorpdfstring{$\Lambda(\vec{x})$}{Λ(x)}}\label{Appendix:Lambda}

In order to define the $\Lambda(\vec{x})$ function, we have to refer to the typical spectra for non-looped interacting polymers. In Ref.~\cite{Foldes2021}, the spectra for polymers throughout the coil-globule transition are studied. For perfect coils, the PSD $\langle X_p^2 \rangle$ follows a single power law $\langle X_p^2 \rangle \sim p^{-(1+2\nu)}$ as a function of $p$, where $\nu$ is the Flory exponent. As the monomer-monomer interaction $\epsilon$ is increased, two power laws are observed: one for the high modes (which roughly corresponds to that of a perfect coil), and one for the low modes. The power law off the low modes goes through a smooth transition from $p^{-(1+2\nu)}$ for perfect coils to $p^0$ for perfect globules. 

Since the even modes $\langle X_{2p}^2 \rangle$ are expected to be the same for a looped and a non-looped polymer, we access the power law of the low modes by fitting the second and fourth mode. By extrapolating to $p=1$, we can then find the expected outcome for a non-looped polymer, and compare it to the actually observed first mode. Calculating this explicitly, we find 
\begin{align*}
        \left[ \frac{ \log\left(\left\langle X_4^2\right\rangle\right) - \log\left(\left\langle X_2^2\right\rangle\right) }{\log(4) - \log(2)} \left( \log(1) - \log(2) \right) + \log\left( \left\langle X_2^2\right\rangle \right) \right] - \log\left(\left\langle X_1^2\right\rangle\right)
        \\
        = 
        -\log\left(\left\langle X_4^2\right\rangle\right) + 2 \log\left(\left\langle X_2^2\right\rangle\right) - \log\left(\left\langle X_1^2\right\rangle\right)
        = \log\left( \frac{ \left\langle X_2^2\right\rangle^2 }{ \left\langle X_1^2 \right\rangle \left\langle X_4^2\right\rangle } \right),
\end{align*}
which is exactly how we defined the log-spectral ratio $\Lambda(x)$ in \Cref{eq:Lambda_def}.

For a random walk $\vec{u}$ of variance $\sigma^2$ and length $N$, we can plug in the spectrum 
\begin{equation}
\langle U_p^2 \rangle = \frac{\sigma^2}{8}\frac{1}{N \sin^2\left( \frac{p\pi}{2N} \right) }    
\label{eq:RW-PSD}
\end{equation}
into the definition of $\Lambda$ to find
\[ \Lambda(\vec{u}) = \log \left( \frac{\cos^2\left(\dfrac{\pi}{N}\right) }{ \cos^2\left(\dfrac{\pi}{2N}\right) } \right) = - \frac{3\pi^2}{4} \frac{1}{N^2} + \mathcal{O}\left( \frac{1}{N^4} \right). \]

For a looped random walk $\vec{l}$, the spectrum
\begin{equation}
    \frac{\left\langle L_p^2\right\rangle }{\left\langle U_p^2 \right\rangle }
    = 
    \begin{cases}
        1 &\text{if } p \textrm{ even} \\
        1 -  2 \left(N \tan\left( \frac{p\pi}{2N}\right) \right)^{-2} &\text{if } p \textrm{ odd}
    \end{cases}
    \label{eq:LoopRW-PSD}
\end{equation}

gives rise to the following log-spectral ratio:
\begin{equation*}
    \Lambda(\vec{\ell}) = \Lambda(\vec{u}) - \log\left( \frac{\left\langle L_1^2 \right\rangle}{\left\langle U_1^2 \right\rangle} \right) = \log\left( \frac{\pi^2}{\pi^2-8} \right) + \mathcal{O}\left(\frac{1}{N^2} \right)
    \approx 1.66 + \mathcal{O}\left(\frac{1}{N^2} \right).
\end{equation*}

\section{Maxima coordinates in the \texorpdfstring{$\Lambda$}{Λ}-plot}\label{Appendix:maxima}

We show how to proceed to determine the loop coordinates $(\iota,\theta)$ from the determination of the maxima $(\iota, \eta)$ in the $\Lambda$-plot. The midpoint of the loop coincides with the first maxima coordinate and can therefore be directly determined. The loop size $\theta$, however, cannot immediately be seen from the $\Lambda$-plot. To relate $\eta$ at the maximum of the $\Lambda$-plot with the actual loop size $\theta$, we perform the following analysis. 

\begin{figure}
    \centering
    \includegraphics[width=\linewidth]{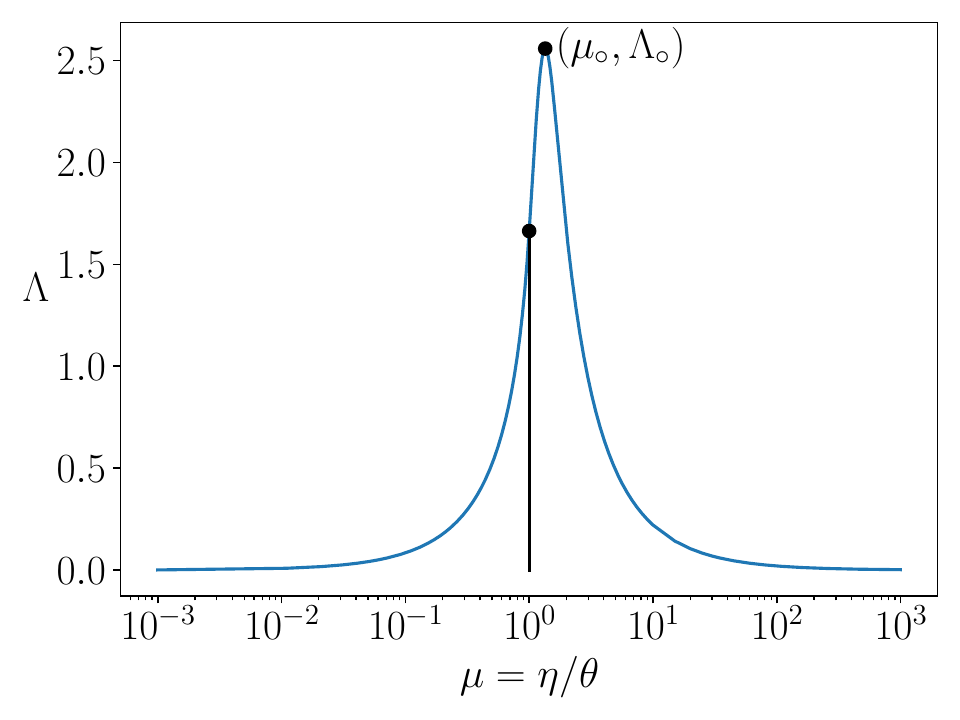}
    \caption{Theoretical midlines of $\Lambda$-plot. Given any fixed $\theta$, and a polymer ensemble, each of fixed size $N$ and with (inner) loop size $\theta$, we can take sub-polymer-ensembles of size $\eta$ centered around the midpoint of the loop. The log-spectral ratio $\Lambda$ for each sub-polymer than varies as given in \Cref{eq:4.13} and as plotted in this figure. This function is smooth everywhere, except at $\mu = \eta/\theta = 1$, where it is only once differentiable. The function has a single maximum at $\mu = \mu_0$ with value $\Lambda_0$ as given in \Cref{eq:value_mu_0,eq:value_lambda_0}. }
    \label{fig:theoretical_midline_rocketplot}
\end{figure}

We take a cross-section of the $\Lambda$-plot for fixed $\iota$, namely the $\iota=\iota_\mathrm{max}$ of the maximum. Hence, only $\eta$ varies. Any given $\eta < \theta$ corresponds to selecting a sub-walk that is contained inside the loop; for $\eta > \theta$, we are selecting the entire loop and some of the adjacent ends. In both cases, the midpoint of the loop is in the middle of the sub-chain. By describing this problem with a 3D random walk, we can explicitly calculate the DCT to find the spectrum of both a partial loop ($\eta < \theta$) and a loop with non-looped ends ($\eta > \theta$). From these spectra, we can calculate the log-spectral ratio $\Lambda$. By writing 
\[ \mu = \eta / \theta,\]
we can discriminate between the two cases with one parameter and we can go to the continuum limit while keeping the ratio $\mu$ fixed. We find then
\begin{equation}\label{eq:4.13}
    \Lambda_0 = \Lambda\big\vert_{\iota=\iota_\mathrm{max}} (\mu)
    = 
    \begin{cases}
        - \log\left( 1 -  \dfrac{8\mu}{\, \, \pi^2} \right) & \mu \leq 1, \\
        - \log\left( 1 - \dfrac{8 \mu }{\pi^2} \sin^2\left( \dfrac{\pi}{2\mu} \right) \right) & \mu > 1.
    \end{cases}
\end{equation} 
Note that this function is once differentiable and has a single maximum. In \Cref{fig:theoretical_midline_rocketplot} we plot \Cref{eq:4.13}.

This analytical expression allows us to understand the relation between $\eta$ and $\theta$. To find the maximum of the function in \Cref{eq:4.13}, we take its derivative and equate it to zero, which yields
\[ \frac{\pi}{\mu} \sin\left( \frac{\pi}{\mu} \right) + \cos\left( \frac{\pi}{\mu} \right) - 1 =0. \]
Manipulating this, we find that finding the position of the maximum is equivalent to solving 
\[ 2x = \tan x  \qquad \text{where } \mu = \frac{1}{x} \frac{\pi}{2} \textrm{ and } x \in \left[ \frac{\pi}{4}, \frac{\pi}{2} \right]. \]
This equation cannot be solved analytically, but a numerical solution gives 
\begin{equation}\label{eq:value_mu_0}
    \mu_0 = 1.34767
\end{equation}
with corresponding value
\begin{equation}\label{eq:value_lambda_0}
    \Lambda\big\vert_{\iota=\iota_\mathrm{max}}( \mu_0) = 2.55882.
\end{equation}

Hence, given the position of a maximum $(\iota, \eta)$, the size of the loop is estimated by $\theta = \eta / \mu_0$.

\section{Log-spectral ratio in mixed populations}\label{Appendix:mixed}

\begin{figure}
    \centering
    \includegraphics[width=\linewidth]{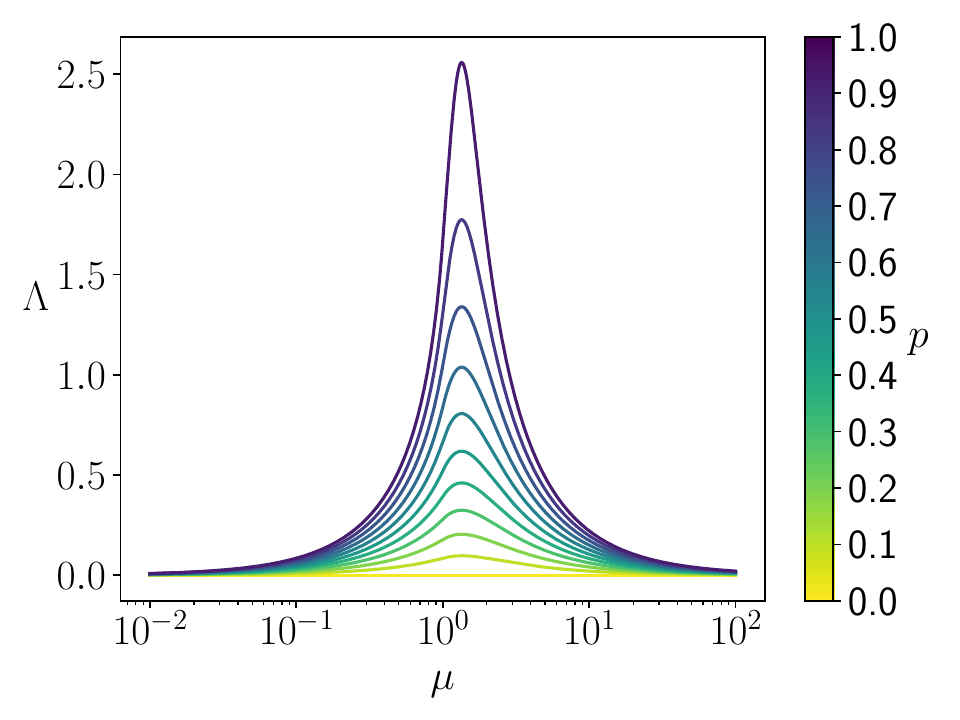}
    \caption{ Theoretical midlines of $\Lambda$-plot for different values of $p$. For $p=1$, the plot of \Cref{fig:theoretical_midline_rocketplot} is recovered. For $p=0$, the log spectral ratio vanishes. For other values of $p$, we plot \Cref{eq:4.17}. The maximum of each function remains at $\mu = \mu_0$ as given in \Cref{eq:value_mu_0}, but the value of the maximum decreases as $p$ decreases. }
    \label{fig:theoretical_midline_rocketplot_mixed_populations}
\end{figure}

Just as in \Cref{Appendix:maxima}, we denote by $\theta$ the size of the loop, and we let $\mu = \eta / \theta$, where $(\iota, \eta)$ are the coordinates used in the $\Lambda$-plot. Suppose $p$ gives the percentage of samples that have a loop in a data set, and hence that $1-p$ gives the percentage of samples that do not have a loop. By linearity,
\[ p \left\langle L_p^2 \right\rangle + (1-p) \left\langle U_p^2 \right\rangle \] is the spectrum for this mixed population, where the $\left\langle L_p^2 \right\rangle$ denotes the spectrum for a uniform population of polymers with the given loop and $\left\langle U_p^2 \right\rangle$ denotes the spectrum of a non-looped population. 
Using this expression within the definition of the log-spectral ratio, we find after going to the continuum limit 
\begin{equation}\label{eq:4.17}    
    \Lambda_\textrm{mixed}(\mu) = 
    \begin{cases}
        - \log\left( 1 - \dfrac{8}{\pi^2} p \mu \right) & \mu \leq 1, \\
        - \log\left( 1 - \dfrac{8}{\pi^2} p \mu \sin^2\left( \dfrac{\pi}{2 \mu} \right)\right) & \mu > 1. \\
    \end{cases}
\end{equation}

This equation relies the intensity of the $\Lambda$ signal to the probability $p$. By taking the equation at the observed maximum, $\Lambda_\textrm{mixed}(\mu_0) = \Lambda_\mathrm{max}$, and by inverting it, we can therefore recover an estimate of the proportion $p$ of samples with loops as
\begin{equation}
\label{eq:proportion_for_lambda_max}
    p =  \dfrac{\pi^2}{8 \mu_0} \left( 1 - e^{-\Lambda_{\text{max}}} \right) \csc^{2}\left( \dfrac{\pi}{2 \mu_0} \right).
\end{equation}

\section{Neural network approach}\label{Appendix:NN}

\begin{table}[tbh]
\caption{Accuracy of trained neural networks for each loop.}\label{table:NN_accuracy}
\begin{tabular}{c|ccccccc}
loop id    & 1 & 2 & 3 & 4 & 5 & 6 & 7 \\ \hline
accuracy [\%]  & 81 & 90 & 95 & 95 & 94 & 92 & 92 
\end{tabular}\vspace{2mm}
\begin{tabular}{c|ccccccc}
loop id    & 8 & 9 & 10 & 11 & 12 & 13 & 14 \\ \hline
accuracy [\%]  & 93 & 83 & 88 & 92 & 92 & 94 & 94 \\
\end{tabular}\vspace{2mm}
\begin{tabular}{c|ccccccc}
loop id  & A1 & A2 & A3 & A4 & A5 & A6 & A7\\ \hline
accuracy [\%]  & 95 & 94 & 95 & 95 & 94 & 93 & 94 \\
\end{tabular}
\end{table}

Neural networks form a class of universal function approximators, meaning that by choosing the appropriate network and giving enough training data, the neural network can---in principle---mimic any function. In this work, we try to approximate the function that takes a polymer and outputs a yes or no answer to the question ‘does this polymer contain a loop?’. In essence, this means we are applying the techniques of logistic regression on higher dimensional input spaces.

To allow the neural network to mimic the above specified function, we need to supply it with sufficient training data. This is data where the correct labels (yes or no) are known, so that the neural network can essentially adapt its fitting parameters to better answer the yes or no question. To avoid over fitting, validation data needs to be supplied as well, and separate test data is required to test the accuracy of the model. Since we will be training on (looped) random walks, we can ourselves generate as many training; validation; and test samples as needed. This is a crucial benefit of this approach.

The necessity of sufficient training data makes it impossible to start from the measurements directly and just start looking for loops. Indeed, many loops can occur together, or intertwined, and all of these possible configurations need many samples to train on. This is why the $\Lambda$-plots developed in this work are used to pinpoint possible locations of loops in the sample which can then, loop by loop, be investigated with a neural network trained to distinguish having one loop or no loops at all. 

The specific networks and inputs we consider, are discussed in \nameref{sec:methods}. Here, we additionally report the accuracy of each trained network in \Cref{table:NN_accuracy}. Note that this accuracy is obtained by studying independent test data, which is (nevertheless) ideal random walk data, and should hence be interpreted carefully. A conclusion we can make is that for the ideal data, the accuracy is around 90 percent, and that the accuracy is lower for shorter loops. This indicates that the neural networks have more trouble separating short random walks and short looped random walks. This is expected as the thermal fluctuations of each single monomer weigh more heavily on the total conformation, when the total number of monomers is small.

\end{appendices}

\end{document}